\documentclass[aps,prl,10pt,twocolumn,floatfix,superscriptaddress,preprintnumbers,shownopacs,showkeys,nofootinbib,notoccite,notitlepage]{revtex4-1}
\usepackage[utf8]{inputenc}
\usepackage[colorlinks=true,citecolor=blue,linkcolor=blue]{hyperref}
\usepackage[normalem]{ulem}
\usepackage{amsmath,amssymb, mathrsfs}
\usepackage{epsfig}
\usepackage{graphicx}               
\usepackage{url}
\usepackage{color}
\usepackage{slashed}
\usepackage{multirow}
\usepackage{placeins}
\usepackage[dvipsnames]{xcolor}
\usepackage{epstopdf}
\usepackage{soul}
\usepackage{tikz}
\usepackage[capitalise, english]{cleveref}
\usepackage{siunitx}
\usepackage{xspace}
\usepackage{booktabs}
\usepackage{makecell}
\usepackage{soul}
\usepackage{tabularx}
\usetikzlibrary{trees}
\usetikzlibrary{decorations.pathmorphing}
\usetikzlibrary{decorations.markings}

\newcommand{\appendixhead}%
{\begin{center}\textbf{\\Appendices\vspace{-0.5cm}}\end{center}}

\newcommand\myshade{80}
\colorlet{mylinkcolor}{ForestGreen}
\colorlet{mycitecolor}{Red}
\colorlet{myurlcolor}{violet}

\hypersetup{
  linkcolor  = mylinkcolor!\myshade!black,
  citecolor  = mycitecolor!\myshade!black,
  urlcolor   = myurlcolor!\myshade!black,
  colorlinks = true
}

\allowdisplaybreaks

\usepackage{tikz,xcolor,hyperref}

\definecolor{lime}{HTML}{A6CE39}
\DeclareRobustCommand{\orcidicon}{\hspace{-1mm}
	\begin{tikzpicture}
	\draw[lime, fill=lime] (0,0) 
	circle [radius=0.16] 
	node[white] {{\fontfamily{qag}\selectfont \tiny \,ID}};
	\draw[white, fill=white] (-0.0525,0.095) 
	circle [radius=0.007];
	\end{tikzpicture}
	\hspace{-3mm}
}

\foreach \x in {A, ..., Z}{\expandafter\xdef\csname orcid\x\endcsname{\noexpand\href{https://orcid.org/\csname orcidauthor\x\endcsname}
			{\noexpand\orcidicon}}
}

\begin{document}

\title{Can LIGO Detect Non-Annihilating Dark Matter?}

\author{Sulagna Bhattacharya\orcidA{}}
\email{sulagna@theory.tifr.res.in}
\affiliation{Tata Institute of Fundamental Research, Homi Bhabha Road, Mumbai 400005, India}

\author{Basudeb Dasgupta\orcidB{}}
\email{bdasgupta@theory.tifr.res.in}
\affiliation{Tata Institute of Fundamental Research, Homi Bhabha Road, Mumbai 400005, India}

\author{Ranjan Laha\orcidC{}}
\email{ranjanlaha@iisc.ac.in}
\affiliation{Centre for High Energy Physics, Indian Institute of Science, C. V. Raman Avenue, Bengaluru 560012, India}

\author{Anupam Ray\orcidD{}}
\email{anupam.ray@berkeley.edu}
\affiliation{Department of Physics, University of California Berkeley, Berkeley, California 94720, USA}
\affiliation{School of Physics and Astronomy, University of Minnesota, Minneapolis, MN 55455, USA}

\date{\today}


\begin{abstract}
Dark matter from the galactic halo can accumulate in neutron stars and transmute them into sub-2.5\,$M_\odot$ black holes if the dark matter particles are heavy, stable, and have interactions with nucleons. We show that non-detection of gravitational waves from mergers of such low-mass black holes can constrain the interactions of non-annihilating dark matter particles with nucleons. 
We find benchmark constraints with LIGO O3 data, viz., $\sigma_{\chi n} \geq {\cal O}(10^{-47})$\,cm$^2$ for bosonic DM with $m_\chi\sim$~PeV (or $m_\chi\sim$~GeV, if they can Bose-condense) and $\geq {\cal O}(10^{-46})$\,cm$^2$ for fermionic DM with $m_\chi \sim 10^3$\,PeV. 
These bounds depend on the priors on DM parameters and on the currently uncertain binary neutron star merger rate density. However, with increased exposure by the end of this decade, LIGO will probe cross-sections that are many orders of magnitude below the neutrino floor and completely test the dark matter solution to missing pulsars in the Galactic center, demonstrating a windfall science-case for gravitational wave detectors as probes of particle dark matter.
\end{abstract}

\maketitle
\preprint{TIFR/TH/23-1, N3AS-23-006}

\emph{Introduction ---} Dark matter (DM) is arguably the most compelling evidence for new physics. Extant searches have placed stringent constraints on non-gravitational interactions of DM in a wide variety of particle physics scenarios~\cite{Zurek:2013wia,	Petraki:2013wwa,Cooley:2022ufh,Boddy:2022knd,Baryakhtar:2022hbu,Carney:2022gse}. However, a simple scenario -- a heavy non-annihilating DM with feeble interactions with the ordinary matter -- remains inadequately tested because of tiny fluxes in terrestrial detectors. 

The leading constraint in this regime arises from the existence of old neutron stars (NSs), which would have imploded to black holes (BHs) due to gradual DM accretion, if DM were to be a heavy non-annihilating particle which interacted with nucleons~\cite{Goldman:1989nd,Gould:1989gw,Bertone:2007ae,deLavallaz:2010wp,McDermott:2011jp,Kouvaris:2010jy,Kouvaris:2011fi,Bell:2013xk,Guver:2012ba,Bramante:2013hn,Bramante:2013nma,Kouvaris:2013kra,Bramante:2014zca,Garani:2018kkd,Kouvaris:2018wnh,Dasgupta:2020dik,Lin:2020zmm,Dasgupta:2020mqg,Goldman:1989nd,Gould:1989gw,Bertone:2007ae,deLavallaz:2010wp,McDermott:2011jp,Kouvaris:2010jy,Kouvaris:2011fi,Bell:2013xk,Guver:2012ba,Bramante:2013hn,Bramante:2013nma,Kouvaris:2013kra,Bramante:2014zca,Fuller:2014rza,Bramante:2017ulk,Garani:2018kkd,Kouvaris:2018wnh,Dasgupta:2020dik,Lin:2020zmm,Dasgupta:2020mqg,Takhistov:2020vxs,Garani:2021gvc,Garani:2022quc,Steigerwald:2022pjo}. More specifically, the strongest constraint in this regime comes from the existence of a Gyr old pulsar close to Earth~\cite{McDermott:2011jp,Garani:2018kkd,Dasgupta:2020dik}; even though NSs in denser parts of the Galaxy are predicted to be more susceptible to DM-induced implosion. This is in part because no old NSs have been detected in the denser inner parts of the Galaxy. In particular, the central parsec of the Galaxy shows a significant deficit of NSs~\cite{Dexter:2013xga}. While there are plausible astrophysical and observational explanations for the observed deficit, it has also led to speculations that the missing pulsars are a hint that NSs near the Galactic center have converted to BHs by accreting heavy non-annihilating DM~\cite{Bramante:2014zca}. This curious situation, coupled with the need to adequately probe heavy non-annihilating DM, demands new ideas.

In this \textit{Letter}, we argue that {gravitational wave} (GW) detectors are a novel and complementary probe of heavy non-annihilating DM interactions with the baryonic matter. 
The key idea is that continued accumulation of DM particles in the NSs leads to anomalously low-mass BHs  in the mass range $\sim$ \mbox{$(1-2.5)\,M_\odot$}, and GWs from such low-mass BH mergers can be searched for by the LIGO-Virgo-KAGRA detector network. Given null detection so far, one already finds an interesting constraint on non-annihilating  DM interactions. This constraint is contingent on the value of the \emph{binary NS} (BNS) merger rate density, which has large uncertainties at present. If it takes the larger values currently allowed, the GW-inferred constraint can be the strongest constraint on DM interactions. With continued data-taking, the existing detectors promise unprecedented sensitivity to non-annihilating  DM interactions, revealing a new windfall science-case for these remarkable detectors.

\emph{Mergers of Low-Mass BHs  ---}
We consider the following sequence of events. A pair of NSs can be born and almost contemporaneously get locked into a binary at time $t_f$. The NSs then accrete DM for a period $\tau_{\rm collapse}$ from the galactic halo, at which point the DM accumulated in their cores collapses to tiny BHs. Then, the tiny BHs take a time $\tau_{\rm swallow}$ to transmute each host NS to a low-mass BH, that we call a \emph{transmuted BH} (TBH)~\cite{Takhistov:2017bpt}. The net transmutation time is $\tau_{\rm trans}=\tau_{\rm collapse}+\tau_{\rm swallow}$. Mergers of these TBH-TBH pairs are detectable at the present time $t_0$ if $t_0-t_f > \tau_{\rm trans}\,$. These timescales are computed in the following.

DM particles that transit through an optically thin NS can get captured due to their collisions with the stellar material. Considering contact interactions of DM with nucleons, one finds a capture rate~\cite{McDermott:2011jp,Garani:2018kkd} 
\begin{align}
	&C = 1.4\times 10^{20}\,{\rm s}^{-1}\,\Big(\tfrac{\rho_{\chi}}{0.4\, \rm{GeV\,cm^{-3}}}\Big) \Big(\tfrac{10^5\, \rm{GeV}}{m_{\chi}}\Big)\Big(\tfrac{\sigma_{\chi n}}{10^{-45}\,\rm{cm^2}}\Big) \nonumber\\
	&\phantom{C = }\times\left(1-\tfrac{1-e^{-A^2}}{A^2}\right)\,\left(\tfrac{v_{\rm{esc}}}{1.9\times10^{5}\rm km\,s^{-1}}\right)^2\left(\tfrac{220\,\rm km\,s^{-1}}{\bar{v}_{\rm gal}}\right)^2\,,
\end{align}
which depends on the ambient DM density $\rho_{\chi}$, the DM mass $m_\chi$, as well as its total interaction cross-section with nucleons $\sigma_{\chi n}$. The factor involving $A^2={6\,m_{\chi}m_n}{v^2_{\rm{esc}}}/{\bar{v}^2_{\rm gal}}{(m_{\chi}-m_n)^2}$ accounts for inefficient momentum transfers at larger $m_\chi$, given NS escape speed $v_{\rm esc}$ and typical DM speeds ${\bar{v}_{\rm gal}}$ in the galaxy. For a typical NS with mass $M_{\rm NS}=1.35\,{M_{\odot}}$ and radius of $R_{\rm NS}=10$\,km, the optical thinness requires $\sigma_{\chi n} \leq 1.3 \times 10^{-45}$\,cm$^2$. For larger cross-sections, the effects of multiple collisions are relevant and it mildly increases the capture rate at larger $m_{\chi}$~\cite{Bramante:2017xlb,Dasgupta:2019juq}.  We  neglect possible self-interactions among the DM particles and nuclear effects in the capture rate~\cite{Anzuini:2021lnv,Bell:2020obw}.

The captured DM, because of the strong gravitational potential of the neutron stars, sinks towards the core, thermalizes, and can collapse to a tiny black hole over a timescale ${\tau_{\rm collapse}} = C^{-1}N^{\rm{BH}}_{\chi}$, where $N^{\rm{BH}}_{\chi}= \max \left[ N^{\rm{self}}_{\chi}, N^{\rm{Cha}}_{\chi}\right]$ denotes the number of DM particles that need to be captured and thermalized to create a nascent BH, cf. refs.~\cite{McDermott:2011jp,Kouvaris:2011fi,Garani:2018kkd,Dasgupta:2020dik}.  $N^{\rm{self}} \sim 1/m^{5/2}_{\chi}$ encodes the Jeans instability criterion, and is determined by  the condition that DM density has to exceed the
baryonic density within the stellar core.  $N^{\rm{Cha}} \sim 1/m^{2}_{\chi}~({\rm or~}1/m^{3}_{\chi})$ denotes the Chandrasekhar limit for bosonic (fermionic) DM, and is set by the effective pressure imbued by quantum mechanics, to prevent this collapse. Detailed numerical estimates, accounting for possible Bose-Einstein condensate (BEC) formation, are reviewed in the Supplemental Material (SM).

The nascent BH, with a very small mass $M_{\rm{BH}} =m_{\chi} N^{\rm{BH}}_{\chi}$,  consumes the NS host over a timescale of $\tau_{\rm{swallow}}= 10^{12}\,{\rm s}\,\left(\tfrac{10^{-16}\,M_{\odot}}{M_{\rm BH}}\right)$~\cite{East:2019dxt, Baumgarte:2021thx,Richards:2021upu,Schnauck:2021hlm}, significantly smaller than stellar lifetimes. Hawking radiation and quantum aspects of accretion can slow down this effect for seed BHs of mass $M_{\rm BH}\gtrsim 10^{-19}\,M_{\odot}$~\cite{Giffin:2021kgb}, providing a maximum DM mass of $\mathcal{O} (10^7)$\,GeV (for bosons) and $\mathcal{O} (10^{10})$\,GeV (for fermions)~\cite{McDermott:2011jp,Kouvaris:2011fi,Garani:2018kkd,Dasgupta:2020dik,Giffin:2021kgb} for transmutation.   The particle DM parameter space that leads to a successful transmutation of NSs is reported in~\cite{Dasgupta:2020mqg}.

The TBH merger rate density~\cite{Dasgupta:2020mqg}
\begin{align}
	R_{\rm{TBH}} &=  \int dr \frac{df}{dr}  \int_{t_*}^{t_0} dt_f  \frac{dR_{\rm BNS}}{dt_f}&  \\
	&\times  \Theta\Big[t_0-t_f-\tau_{\rm{trans}}\left [m_\chi,\sigma_{\chi n},\rho_{\textrm{ext}}(r,t_0)\right] \Big]\,,&\nonumber
	\label{eq:TBH}
\end{align}
is a fraction of the BNS merger rate density ($R_{\rm BNS}$) that one would have \emph{if there were no transmutations}, depending on the DM properties through $\tau_{\rm trans}$, and on astrophysical conditions. We assume a uniform 1d distribution $df/dr$ of progenitor BNSs in Milky-Way like galaxies, where $r\in(0.01,0.1)$\,kpc denotes the galactocentric distance. This affects the background DM density $\rho_{\rm ext}$ experienced by the progenitors, which we take to have a Navarro-Frenk-White profile $\rho_{\rm{ext}}[r,t_0]=\rho_{\rm{ext}}[r] = \rho_s/\left((r/r_s)(1+r/r_s)^2\right)$~\cite{Navarro:1995iw,Navarro:1996gj}, where $\rho_s=0.47\,\mathrm{GeV\,cm^{-3}}$ and $r_s=14.5\,\mathrm{kpc}$ for a Milky-Way like galaxy. Note that we do not consider the time evolution of the ambient DM density, and use its current value, i.e., $\rho_{\rm{ext}}(z=0)$, in order to be conservative.  The lower-limit of the $t_f$ integral, $t_*$, corresponds to $z_*=10$, taken as the the earliest formation time and $dR_{\rm BNS}/dt_f$ is the rate-density of progenitor mergers for a given formation time~\cite{Taylor:2012db}. The latter is proportional to the star formation rate $d\rho_*/dt$ (for which we take the Madau-Dickinson model~~\cite{Madau:2014bja}), the fraction of stellar mass in binaries $\lambda\approx10^{-5}$~\cite{Taylor:2012db,OShaughnessy:2007brt, OShaughnessy:2009szr}, and their merger time distribution at present time proportional to $(t_0-t_f)^{-1}$~\cite{Taylor:2012db}. Only the shape of ${dR_{\rm BNS}}/{dt_f}$ is an independent assumption because the overall normalization $R_{\rm BNS}$ is taken to be a free parameter in the range $(10-1700)\,{\rm Gpc}^{-3}{\rm yr}^{-1}$ favored by recent LIGO observations~\cite{LIGOScientific:2021psn}. In addition to the above, we take $R_{\rm{NS}}=10$\,km, $T_{\rm{core}} = 2.1 \times 10^6$\,K, and a monochromatic mass distribution of the progenitors centered at 1.35\,$M_{\odot}$ for computing $\tau_{\rm trans}$. The dependence of $R_{\rm{TBH}}$ on various model assumptions is studied in the SM which additionally includes Refs.~\cite{Acevedo:2020gro,Ray:2023auh,Boudaud:2014qra,1980ApJS...44...73B,Steidel:1998ja,Porciani:2000ag,2010MNRAS.402..371B,Dominik:2012kk,Vitale:2018yhm,Santoliquido:2020axb,Mukherjee:2021qam,Steigman:2012nb,Barsanti:2021ydd,Joglekar:2019vzy,Bell:2020lmm,Nunes:2021jsk}.

\begin{figure*}[t]
	\includegraphics[width=0.32\textwidth]{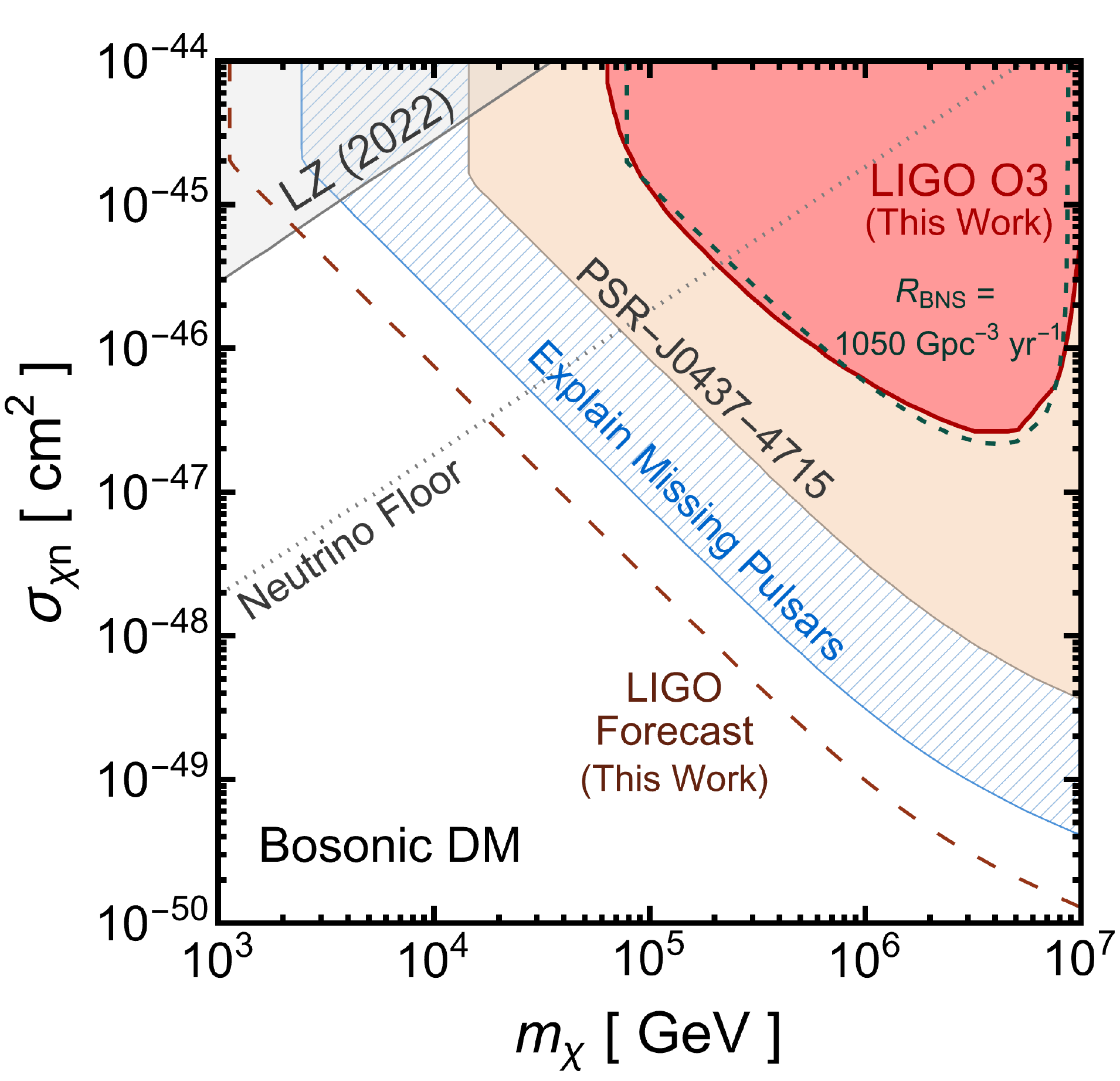}
	\includegraphics[width=0.32\textwidth]{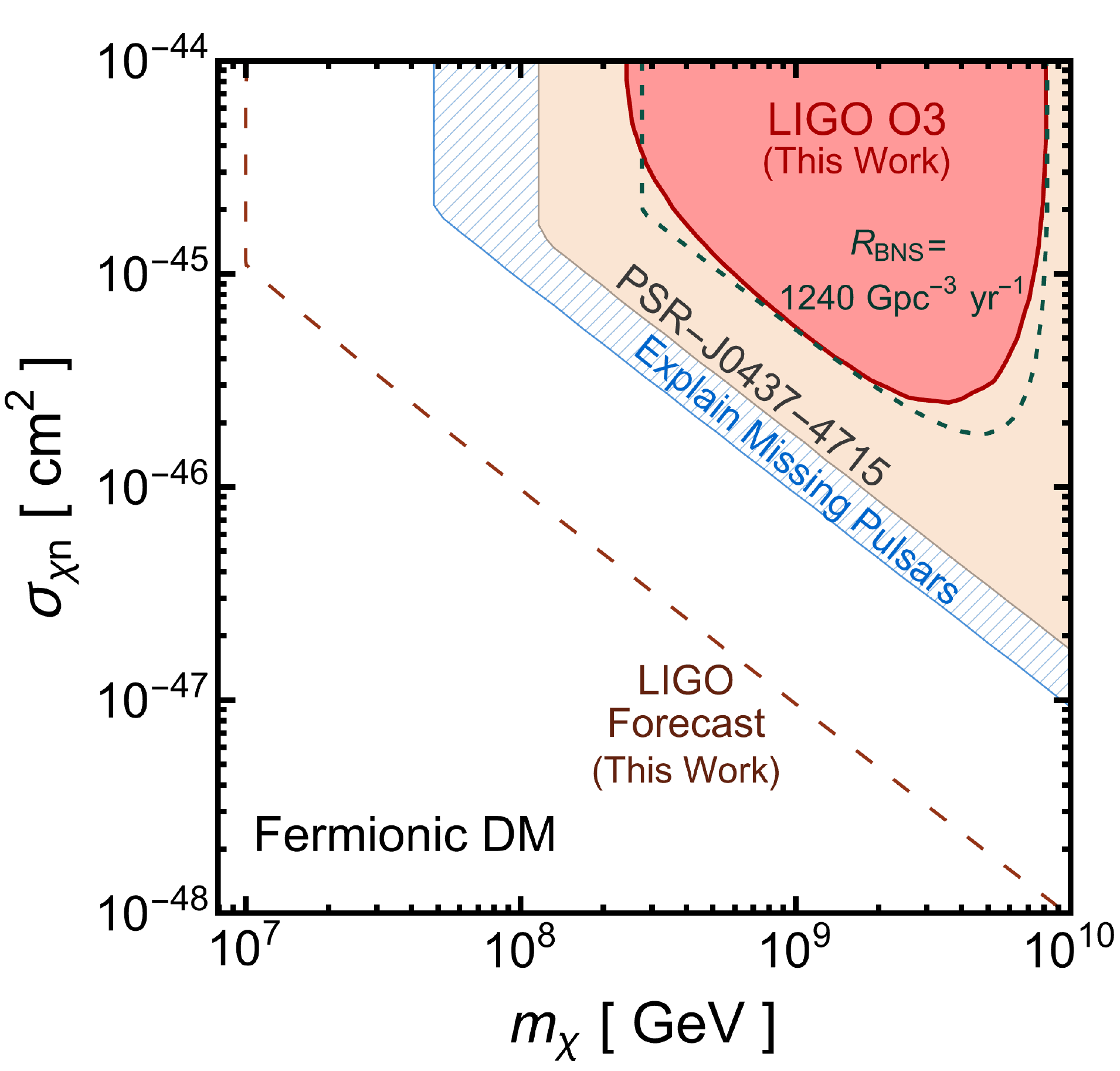}
	\includegraphics[width=0.315\textwidth]{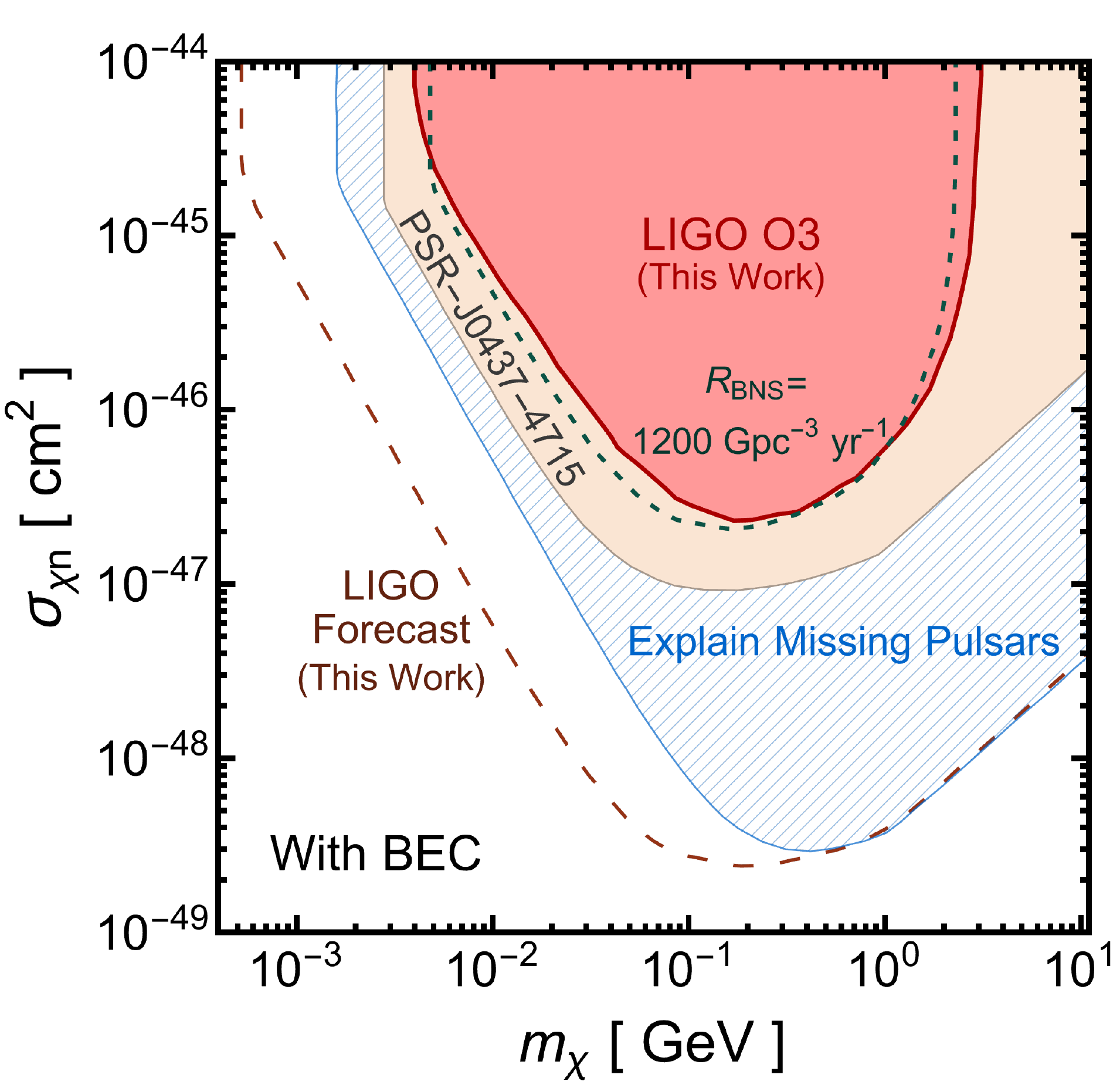}
	\caption{Gravitational wave constraints on bosonic (left panel:\,without BEC, right panel:\,with BEC) and fermionic (middle panel) non-annihilating dark matter interactions with nucleons. These constraints apply to both spin-(in)dependent interactions as DM-neutron scattering is considered. Non-detection of BBH mergers by the LIGO O3 low-mass BH search (MBTA pipeline)~\cite{LIGOScientific:2022hai} disfavors the pink shaded regions, as per our nominally 90\% credible Bayesian limit obtained by marginalizing over $R_{\rm BNS}\in\,$(10\,-\,1700)\,Gpc$^{-3}$yr$^{-1}$. A frequentist 90\% confidence upper limit, obtained by assuming $R_{\rm BNS} = 1050\, ({\rm or}~1240,~{\rm or}~1200)\, \mathrm{Gpc^{-3}\,yr^{-1}}$, shown with the green dashed line, roughly matches the corresponding Bayesian limits. The brown dashed line is a forecasted 90\% confidence frequentist upper limit obtainable with $50$ times the current exposure $\langle VT\rangle$ and marginalizing over currently allowed range of $R_{\rm BNS}$. The leading constraint from terrestrial experiments is shown as \mbox{``LZ (2022)''} (spin-independent) in the left panel~\cite{LZ:2022lsv}. The hatched blue region, labeled by ``Explain Missing Pulsars'', shows parameter space that would address the missing pulsar problem by invoking NS transmutation to BHs via DM accretion, without being in conflict with the existence of known pulsars, specifically PSR-J0437-4715, that disfavors the beige shaded region towards top-right. We also show the neutrino floor for the direct detection experiments, below which potential discovery of a DM signal is hindered by neutrino backgrounds~\cite{OHare:2021utq}; for fermionic DM the neutrino floor is above the range of cross-sections shown. Note that the ranges of  mass and cross-section shown in the three panels are different.}
	\label{gwlimit}
\end{figure*}

\emph{Data and Statistics  ---} We estimate the TBH merger rate density for a chirp mass bin
\begin{align}
R_{{\rm TBH},i} = p_i \times R_{\rm{TBH}}\,,
\end{align}
where $p_i$ is the probability that the progenitor BNS has chirp mass $m_c= (m_1 m_2)^{3/5}/(m_1+m_2)^{1/5}$ in the $i^{th}$ bin, given the probability distributions of $m_{1,2}$. The component NS masses are $m_2<m_1$ by convention, with asymmetry parameter $q=m_2/m_1 <1$. The masses $m_{1,2}$ are approximately Gaussian distributed between 1.08\,$M_{\odot}$ and 1.57\,$M_{\odot}$, with mean $\approx 1.35\,M_{\odot}$ and standard deviation $\approx0.09\,M_{\odot}$, as inferred from a large astrophysical sample~\cite{Ozel:2016oaf}. The TBH-TBH mergers are thus predicted around the chirp mass $m_c\approx1.15\,M_{\odot}$. The NS mass distribution predicts $q_{\rm min}=0.69$, consistent with the LIGO search criterion of $q>0.1$~\cite{LIGOScientific:2022hai}. 

 Given the non-detection of low-mass BBHs in the LIGO O3 data ~\cite{LIGOScientific:2022hai}, it is reasonable to assume a Poisson distribution for the event counts in each chirp mass bin. The binned rate density $R_{\textrm{TBH},i}$ depends on the model parameters $\bar{\theta}=\{m_\chi,\, \sigma_{\chi n},\,R_{\rm BNS}\}$. With a surveyed volume-time $\langle VT \rangle_i$, called {exposure} hereafter, the likelihood for parameters $\bar\theta$ is 
\begin{align}
	{L}_i=\exp\big[-R_{\mathrm{TBH},i} \times \langle VT \rangle_i\big]\,,
\end{align}
where we use the $\langle VT \rangle_i$ provided by LIGO (MBTA pipeline) for its third observing run~\cite{LIGOScientific:2022hai}.

We will derive both Bayesian as well as frequentist constraints on the DM parameters. With current data, owing to the uncertainty on $R_{\rm BNS}$, the frequentist limits are not constraining and we show only the Bayesian limits. With more exposure, we find interesting sensitivities without any priors on DM and show frequentist forecasts.

For the Bayesian limits, the posterior for the parameters $\bar{\theta}$ is given by ${P} \big[\bar{\theta}\,\big]\propto\prod_{i} {L}_i\big[\bar{\theta}\,\big]	\times \pi\big[\bar\theta\,\big]\,.$ We assume log-uniform priors on \mbox{$m_{\chi}\,\in\,\left(10^4,\,10^8\right)$\,GeV} for bosonic DM without BEC formation, \mbox{$m_{\chi}\,\in\,\left(10^{-3},\,10^3\right)$\,GeV}\, with BEC, and \mbox{$m_{\chi}\in\left(10^8,\,10^{11}\right)$\,GeV} for fermionic DM, and log-uniform priors on $\sigma_{\chi n} \in \left(10^{-50},\,10^{-44}\right)$\,cm$^2$ for bosonic DM without BEC formation, $\sigma_{\chi n} \in \left(10^{-49},\,10^{-44}\right)$\,cm$^2$ with BEC, and $\sigma_{\chi n} \in \left(10^{-48},\,10^{-44}\right)$\,cm$^2$ for fermionic DM. The ranges for $m_\chi$ and $\sigma_{\chi n}$ are chosen to be somewhat larger than the parameter space where transmutation is possible ($\tau_{\rm trans}<10\, \mathrm{Gyr}$). For smaller cross-sections or masses outside the above ranges, the parameters will not be excluded by the LIGO data. For larger cross-sections the likelihood becomes small, so that the exclusion contour is somewhat sensitive to the choice of the upper-boundary of the prior on $\sigma_{\chi n}$ whenever we obtain a nontrivial constraint. We take a uniform prior on $R_{\rm{BNS}} \in (10,\,1700)$\,\mbox{Gpc$^{-3}$ yr$^{-1}$}~\cite{LIGOScientific:2021psn}.
We sample the 3d posterior distribution of the parameters by using  the {\tt emcee} Markov Chain Monte Carlo sampler~\cite{2013PASP..125..306F}, and marginalize the posterior over the additional parameter $R_{\rm{BNS}}$ to find the marginal 2d posterior of $\left\{m_{\chi},\sigma_{\chi n} \right\}$. We then identify the minimal region of the $m_{\chi}-\sigma_{\chi n}$ plane that contains 90\% of the sampled points to present a 90\% credible constraint in the $m_{\chi}-\sigma_{\chi n}$ plane. 

We obtain frequentist limits using the likelihood $L=\exp[-\mu]$, with $\mu = R_{\rm TBH}\langle VT \rangle$ obtained by assuming a fixed value of $R_{\rm BNS}$. For simplicity, here we approximate that all likelihood is in a chirp mass bin around $m_c=1.15\,M_\odot$. Alternatively, to get hybrid-frequentist limits we use the marginal likelihood
\begin{align}
	{L}_{\rm m}=\frac{e^{-\kappa_{\rm min}\mu} - e^{-\kappa_{\rm max}\mu}}{\mu\,\log[\kappa_{\rm max}/\kappa_{\rm min}]}\,,
		\label{eq:mu90}
\end{align}
obtained by writing $R_{\rm BNS}=\kappa \times 1000$\,{Gpc$^{-3}$yr$^{-1}$ in the likelihood $L$, and averaging over the nuisance parameter $\kappa$ with a uniform prior in $(\kappa_{\rm min},\,\kappa_{\rm max})$. Upper limits on $\mu$ (at 90\% confidence) are then obtained  by setting $\int_{\mu_{90}}^{\infty} d\mu L_{(\rm m)}=0.1$. If $R_{\rm BNS}$ were not uncertain, i.e. $\kappa$ were fixed, there would be no nuisance parameter. In this case, for a null-detection described by a Poisson process without background, the Bayesian and frequentist 90\% upper limits on the expected number of signal events coincide at $2.303$. We will use this to compare our Bayesian constraints with related frequentist limits. 

\emph{GW Limits on DM Parameters ---} In Fig.\,\ref{gwlimit}, the pink shaded regions labeled ``LIGO O3'' show the 90\% credibility disfavored regions of the marginal 2d posteriors of $\{m_{\chi},\,\sigma_{\chi n}\}$ for bosonic, fermionic, and BEC-forming DM.  We find an upper limit of $\sigma_{\chi n}<2.5\times10^{-47}{\rm cm}^2$ for $m_\chi=5$\,PeV (or $0.2$\,GeV) bosonic dark matter without (with) BEC formation, respectively, weakening as $\sim 1/m_\chi ^{3/2}$ (or $\sim 1/m_\chi ^{2}$) at smaller masses up to 0.06 PeV  (or 4\,MeV). For fermionic dark matter, $\sigma_{\chi n}<2.4\times10^{-46}{\rm cm}^2$ for $m_\chi=3.6 \times 10^3$\,PeV, weakening as $\sim1/m_\chi$ up to 240 PeV. These limits are roughly comparable to a lower limit on $\tau_{\rm trans} \leq 0.4\,\rm{Gyr}~({\rm or}~0.3\,\rm{Gyr})$ for  bosonic DM without (with) BEC formation and $\tau_{\rm trans} \leq 3\,\rm Gyr$ for fermionic DM for $\rho_{\chi} =0.4\, \rm{GeV\,cm^{-3}}$. 

The dark green curves labeled by ``$R_{\rm BNS} = 1050$ (or 1240, or 1200)\,$\mathrm{Gpc^{-3}\,yr^{-1}}$'' are frequentist 90\% upper limits obtained by assuming a fixed value of $R_{\rm BNS}$ as noted. Our 90\% credible Bayesian limits are numerically similar to these, allowing us to interpret these constraints in relation to each other. If $R_{\rm BNS}=10\,\rm Gpc^{-3}\,yr^{-1}$, with current data we find no 90\% frequentist constraint on the DM parameter space. The minimum values of $R_{\rm BNS}$ for which current data can start ruling out some of the DM parameter space in a frequentist analysis are approximately $900\,({\rm or}~980,\,{\rm or}~1110)\,\rm Gpc^{-3}\,yr^{-1}$, for bosonic DM without BEC formation, with BEC, and fermionic DM, respectively. We also ask, what is the hybrid-frequentist constraint that exactly mimics our Bayesian analysis, but without having to assume any priors on $m_\chi$ and $\sigma_{\chi n}$. For bosonic DM without BEC formation, using the range $\kappa\in(0.01,1.7)$, the 90\% hybrid upper limit gives $\mu_{90}\approx54$. We recall that our Bayesian constraint is comparable to a 90\% frequentist upper limit assuming $R_{\rm BNS}=1050\,{\rm Gpc}^{-3}{\rm yr}^{-1}$, which in turn is equivalent to taking the limits $\kappa_{\rm max, min}\to 1.05$, for which the 90\% hybrid upper limit gives $\mu_{90}\approx2.2$. The numerical value of $\mu_{90}$ for our hybrid analysis is therefore approximately $54/2.2\approx25$ times larger than for our benchmark Bayesian upper limit. For the case of bosonic DM with BEC formation and fermionic DM we find that our Bayesian limits are nominally stronger by factors of 28 and 29, respectively, compared to the hybrid limits. This is ascribable to the priors on DM parameters.   

In  Fig.\,\ref{gwlimit}, we also show the leading constraint from underground direct detection experiments~\cite{LZ:2022lsv}, in the left panel as a shaded region labeled ``LZ (2022)'', as well as an exclusion limit from the existence of the pulsar PSR-J0437-4715~\cite{McDermott:2011jp,Garani:2018kkd} as a shaded region. This particular pulsar, because of its relatively low core temperature and long lifetime provides the most stringent constraint on weakly interacting heavy non-annihilating DM. Apart from that, because of its close proximity, the ambient DM density and the surface temperature have been measured with small uncertainties, indicating the robustness of this constraint. Our current constraint, inferred from the existing LIGO data, is weaker than the PSR-J0437-4715 constraint. However, because of the entirely different systematics of GW detection as opposed to radio searches for pulsars, it is complementary and it has the potential to set the leading constraint with the upcoming GW observations. In  Fig.\,\ref{gwlimit}, the blue-hatched region shows the DM parameter space that can putatively explain the scarcity of old pulsars in the central parsec of our Galaxy; it corresponds to DM parameters that can transmute all the 30 Myr old pulsars that are within 10 arc-minutes of the Galactic Center. 

The  curves labeled ``LIGO Forecast'' are forecasted hybrid-frequentist upper limits (90\% confidence; marginalized over $R_{\rm BNS}\in (10,\,1700)$\,Gpc$^{-3}$ yr$^{-1}$~\cite{LIGOScientific:2021psn}) that can be obtained in the future if the exposure $\langle VT\rangle$ grows to 50 times the current exposure, as may be possible by the end of this decade~\cite{LVKObs}. Conditionally, if $R_{\rm{BNS}} \gtrsim 28$ $\,\rm{Gpc^{-3}\,yr^{-1}}$, our proposed method can supersede the EM-inferred constraints on non-annihilating DM interactions assuming 50 times more exposure than the current LIGO-O3. With future detectors~\cite{Maggiore:2019uih,Evans:2021gyd}, the sensitivity can improve by several orders of magnitude (see SM for estimates). It is evident that  the forecasted LIGO sensitivity can completely test the DM solution to the missing pulsar problem, and provides perhaps a unique way to probe DM-nucleon cross-sections well below the neutrino floor.

\emph{Summary {\&} Outlook ---} We have argued that non-detection of GWs from mergers of low-mass BBHs can be used to probe the particle nature of DM. Specifically, we use null-detection of such events until the O3 run of the LIGO-Virgo-KAGRA collaboration to infer constraints on interactions of heavy non-annihilating DM with nucleons. Our benchmark constraints disfavor $\sigma_{\chi n} \geq {\cal O}(10^{-47})$\,cm$^2$ for bosonic DM with  PeV-scale mass if no BEC forms, and with GeV-scale if a BEC can form. We find $\sigma_{\chi n} \geq {\cal O}(10^{-46})$\,cm$^2$ for $10^3$-PeV-scale fermionic DM. We note that, the same low-mass BBH searches have recently been used to probe primordial BHs as DM~\cite{LIGOScientific:2005fbz,LIGOScientific:2007npa,LIGOScientific:2018glc,LIGOScientific:2019kan,LIGOScientific:2021job,LIGOScientific:2022hai,Nitz:2021mzz,Nitz:2021vqh,Nitz:2022ltl,Phukon:2021cus} and an atomic DM model~\cite{LIGOScientific:2021job,LIGOScientific:2022hai,Singh:2020wiq}, and this is the \textit{first} attempt to
demonstrate that it also sheds light on $\sigma_{\chi n}$ quite generically for weakly interacting non-annihilating DM.

The presented constraint is sensitive to the uncertainty in the BNS merger rate density and priors on DM parameters. Current LIGO data suggests a broad range for $R_{\rm BNS}\in\,$(10\,-\,1700)\,Gpc$^{-3}$yr$^{-1}$~\cite{LIGOScientific:2021psn}. With current data, the frequentist limits are not constraining unless $R_{\rm BNS}\geq 900$\,Gpc$^{-3}$yr$^{-1}$. On the other hand, if $R_{\rm BNS}\approx$ 1700\,Gpc$^{-3}$yr$^{-1}$, at the upper end of the currently allowed range, GW detectors already provide leading sensitivity to interactions of DM with nucleons. The constraints are modestly sensitive to other astrophysical inputs, mainly the DM density profiles in galactic halos that affect $R_{\rm TBH}$ with a nontrivial $m_\chi$-dependence. Uncertainties in BNS merger time delay distributions, star formation rate, etc., mainly lead to $~50\%$ level normalization uncertainties that are subsumed in the larger uncertainty on $R_{\rm BNS}$.  Uncertainties on the NS properties can cause a small $\sim20\%$ level change. New particle physics such as self-interactions of DM or due to phases of NS matter could be important, but are out of our scope here.

In the future, if there are detections of anomalously low-mass BBHs, it will be important to check if other source-classes could fake a TBH-like signal. Besides novel objects such as primordial BHs, it is plausible that a fraction of BNSs may get incorrectly classified as low-mass BBHs. This can be mitigated if tidal deformation in the events is measured reliably and precisely~\cite{Singh:2022wvw}. In such a case, one would search for TBH events as a signal over the estimated background due to BNS events that were incorrectly classified as BBH events. For null detection, assuming a zero background gives conservative constraints on DM parameters. We anticipate that the sensitivity to TBH mergers can be improved with a more detailed analysis of LIGO data. It is also expected to have a distinctive redshift dependence~\cite{Dasgupta:2020mqg}.

Encouragingly, because of the planned upgrades of the LIGO-Virgo-KAGRA detectors and continued data-taking, one expects spectacular sensitivity to DM parameter space by the end of this decade. We find this to be possible without assuming any priors on DM parameters. GW detectors may be able to look for non-annihilating DM that is much heavier and much more weakly interacting than will be possible using any other probe, covering the entire parameter space that explains the missing pulsars, and going well below the neutrino floor.

\emph{Acknowledgments ---} We thank Joseph Bramante, Rishi Khatri, Girish Kulkarni, Shikhar Mittal, Maxim Pospelov, Nirmal Raj, and Joe Silk for useful discussions. BD is supported by the Dept.~of Atomic Energy~(Govt.~of~India) research project RTI 4002, the Dept.~of Science and Technology~(Govt.~of~India) through a Swarnajayanti {Fellowship}, and by the Max-Planck-Gesellschaft through a {Max Planck Partner Group}. RL acknowledges financial support from the Infosys foundation (Bangalore), institute start-up funds, and Dept. of Science and Technology (Govt. of India) for the grant SRG/2022/001125. AR acknowledges support from the National Science Foundation (Grant No. PHY-2020275) and to the Heising-Simons Foundation (Grant 2017-228).

\bibliographystyle{JHEP}
\bibliography{ref.bib}

\providecommand{\href}[2]{#2}\begingroup\raggedright\begin{thebibliography}{10}

\bibitem{Zurek:2013wia}
K.~M. Zurek, \emph{{Asymmetric Dark Matter: Theories, Signatures, and
  Constraints}},
  \href{https://doi.org/10.1016/j.physrep.2013.12.001}{\emph{Phys. Rept.}
  {\bfseries 537} (2014) 91} [\href{https://arxiv.org/abs/1308.0338}{{\ttfamily
  1308.0338}}].

\bibitem{Petraki:2013wwa}
K.~Petraki and R.~R. Volkas, \emph{{Review of asymmetric dark matter}},
  \href{https://doi.org/10.1142/S0217751X13300287}{\emph{Int. J. Mod. Phys. A}
  {\bfseries 28} (2013) 1330028}
  [\href{https://arxiv.org/abs/1305.4939}{{\ttfamily 1305.4939}}].

\bibitem{Cooley:2022ufh}
J.~Cooley et~al., \emph{{Report of the Topical Group on Particle Dark Matter
  for Snowmass 2021}},  \href{https://arxiv.org/abs/2209.07426}{{\ttfamily
  2209.07426}}.

\bibitem{Boddy:2022knd}
K.~K. Boddy et~al., \emph{{Snowmass2021 theory frontier white paper:
  Astrophysical and cosmological probes of dark matter}},
  \href{https://doi.org/10.1016/j.jheap.2022.06.005}{\emph{JHEAp} {\bfseries
  35} (2022) 112} [\href{https://arxiv.org/abs/2203.06380}{{\ttfamily
  2203.06380}}].

\bibitem{Baryakhtar:2022hbu}
M.~Baryakhtar et~al., \emph{{Dark Matter In Extreme Astrophysical
  Environments}},  in \emph{{2022 Snowmass Summer Study}}, 3, 2022,
  \href{https://arxiv.org/abs/2203.07984}{{\ttfamily 2203.07984}}.

\bibitem{Carney:2022gse}
D.~Carney et~al., \emph{{Snowmass2021 Cosmic Frontier White Paper: Ultraheavy
  particle dark matter}},  \href{https://arxiv.org/abs/2203.06508}{{\ttfamily
  2203.06508}}.

\bibitem{Goldman:1989nd}
I.~Goldman and S.~Nussinov, \emph{{Weakly Interacting Massive Particles and
  Neutron Stars}}, \href{https://doi.org/10.1103/PhysRevD.40.3221}{\emph{Phys.
  Rev. D} {\bfseries 40} (1989) 3221}.

\bibitem{Gould:1989gw}
A.~Gould, B.~T. Draine, R.~W. Romani and S.~Nussinov, \emph{{Neutron Stars:
  Graveyard of Charged Dark Matter}},
  \href{https://doi.org/10.1016/0370-2693(90)91745-W}{\emph{Phys. Lett. B}
  {\bfseries 238} (1990) 337}.

\bibitem{Bertone:2007ae}
G.~Bertone and M.~Fairbairn, \emph{{Compact Stars as Dark Matter Probes}},
  \href{https://doi.org/10.1103/PhysRevD.77.043515}{\emph{Phys. Rev. D}
  {\bfseries 77} (2008) 043515}
  [\href{https://arxiv.org/abs/0709.1485}{{\ttfamily 0709.1485}}].

\bibitem{deLavallaz:2010wp}
A.~de~Lavallaz and M.~Fairbairn, \emph{{Neutron Stars as Dark Matter Probes}},
  \href{https://doi.org/10.1103/PhysRevD.81.123521}{\emph{Phys. Rev. D}
  {\bfseries 81} (2010) 123521}
  [\href{https://arxiv.org/abs/1004.0629}{{\ttfamily 1004.0629}}].

\bibitem{McDermott:2011jp}
S.~D. McDermott, H.-B. Yu and K.~M. Zurek, \emph{{Constraints on Scalar
  Asymmetric Dark Matter from Black Hole Formation in Neutron Stars}},
  \href{https://doi.org/10.1103/PhysRevD.85.023519}{\emph{Phys. Rev. D}
  {\bfseries 85} (2012) 023519}
  [\href{https://arxiv.org/abs/1103.5472}{{\ttfamily 1103.5472}}].

\bibitem{Kouvaris:2010jy}
C.~Kouvaris and P.~Tinyakov, \emph{{Constraining Asymmetric Dark Matter through
  observations of compact stars}},
  \href{https://doi.org/10.1103/PhysRevD.83.083512}{\emph{Phys. Rev. D}
  {\bfseries 83} (2011) 083512}
  [\href{https://arxiv.org/abs/1012.2039}{{\ttfamily 1012.2039}}].

\bibitem{Kouvaris:2011fi}
C.~Kouvaris and P.~Tinyakov, \emph{{Excluding Light Asymmetric Bosonic Dark
  Matter}}, \href{https://doi.org/10.1103/PhysRevLett.107.091301}{\emph{Phys.
  Rev. Lett.} {\bfseries 107} (2011) 091301}
  [\href{https://arxiv.org/abs/1104.0382}{{\ttfamily 1104.0382}}].

\bibitem{Bell:2013xk}
N.~F. Bell, A.~Melatos and K.~Petraki, \emph{{Realistic neutron star
  constraints on bosonic asymmetric dark matter}},
  \href{https://doi.org/10.1103/PhysRevD.87.123507}{\emph{Phys. Rev. D}
  {\bfseries 87} (2013) 123507}
  [\href{https://arxiv.org/abs/1301.6811}{{\ttfamily 1301.6811}}].

\bibitem{Guver:2012ba}
T.~G\"uver, A.~E. Erkoca, M.~Hall~Reno and I.~Sarcevic, \emph{{On the capture
  of dark matter by neutron stars}},
  \href{https://doi.org/10.1088/1475-7516/2014/05/013}{\emph{JCAP} {\bfseries
  05} (2014) 013} [\href{https://arxiv.org/abs/1201.2400}{{\ttfamily
  1201.2400}}].

\bibitem{Bramante:2013hn}
J.~Bramante, K.~Fukushima and J.~Kumar, \emph{{Constraints on bosonic dark
  matter from observation of old neutron stars}},
  \href{https://doi.org/10.1103/PhysRevD.87.055012}{\emph{Phys. Rev. D}
  {\bfseries 87} (2013) 055012}
  [\href{https://arxiv.org/abs/1301.0036}{{\ttfamily 1301.0036}}].

\bibitem{Bramante:2013nma}
J.~Bramante, K.~Fukushima, J.~Kumar and E.~Stopnitzky, \emph{{Bounds on
  self-interacting fermion dark matter from observations of old neutron
  stars}}, \href{https://doi.org/10.1103/PhysRevD.89.015010}{\emph{Phys. Rev.
  D} {\bfseries 89} (2014) 015010}
  [\href{https://arxiv.org/abs/1310.3509}{{\ttfamily 1310.3509}}].

\bibitem{Kouvaris:2013kra}
C.~Kouvaris and P.~Tinyakov, \emph{{Growth of Black Holes in the interior of
  Rotating Neutron Stars}},
  \href{https://doi.org/10.1103/PhysRevD.90.043512}{\emph{Phys. Rev. D}
  {\bfseries 90} (2014) 043512}
  [\href{https://arxiv.org/abs/1312.3764}{{\ttfamily 1312.3764}}].

\bibitem{Bramante:2014zca}
J.~Bramante and T.~Linden, \emph{{Detecting Dark Matter with Imploding Pulsars
  in the Galactic Center}},
  \href{https://doi.org/10.1103/PhysRevLett.113.191301}{\emph{Phys. Rev. Lett.}
  {\bfseries 113} (2014) 191301}
  [\href{https://arxiv.org/abs/1405.1031}{{\ttfamily 1405.1031}}].

\bibitem{Garani:2018kkd}
R.~Garani, Y.~Genolini and T.~Hambye, \emph{{New Analysis of Neutron Star
  Constraints on Asymmetric Dark Matter}},
  \href{https://doi.org/10.1088/1475-7516/2019/05/035}{\emph{JCAP} {\bfseries
  05} (2019) 035} [\href{https://arxiv.org/abs/1812.08773}{{\ttfamily
  1812.08773}}].

\bibitem{Kouvaris:2018wnh}
C.~Kouvaris, P.~Tinyakov and M.~H.~G. Tytgat, \emph{{NonPrimordial Solar Mass
  Black Holes}},
  \href{https://doi.org/10.1103/PhysRevLett.121.221102}{\emph{Phys. Rev. Lett.}
  {\bfseries 121} (2018) 221102}
  [\href{https://arxiv.org/abs/1804.06740}{{\ttfamily 1804.06740}}].

\bibitem{Dasgupta:2020dik}
B.~Dasgupta, A.~Gupta and A.~Ray, \emph{{Dark matter capture in celestial
  objects: light mediators, self-interactions, and complementarity with direct
  detection}}, \href{https://doi.org/10.1088/1475-7516/2020/10/023}{\emph{JCAP}
  {\bfseries 10} (2020) 023}
  [\href{https://arxiv.org/abs/2006.10773}{{\ttfamily 2006.10773}}].

\bibitem{Lin:2020zmm}
G.-L. Lin and Y.-H. Lin, \emph{{Analysis on the black hole formations inside
  old neutron stars by isospin-violating dark matter with self-interaction}},
  \href{https://doi.org/10.1088/1475-7516/2020/08/022}{\emph{JCAP} {\bfseries
  08} (2020) 022} [\href{https://arxiv.org/abs/2004.05312}{{\ttfamily
  2004.05312}}].

\bibitem{Dasgupta:2020mqg}
B.~Dasgupta, R.~Laha and A.~Ray, \emph{{Low Mass Black Holes from Dark Core
  Collapse}}, \href{https://doi.org/10.1103/PhysRevLett.126.141105}{\emph{Phys.
  Rev. Lett.} {\bfseries 126} (2021) 141105}
  [\href{https://arxiv.org/abs/2009.01825}{{\ttfamily 2009.01825}}].

\bibitem{Fuller:2014rza}
J.~Fuller and C.~Ott, \emph{{Dark Matter-induced Collapse of Neutron Stars: A
  Possible Link Between Fast Radio Bursts and the Missing Pulsar Problem}},
  \href{https://doi.org/10.1093/mnrasl/slv049}{\emph{Mon. Not. Roy. Astron.
  Soc.} {\bfseries 450} (2015) L71}
  [\href{https://arxiv.org/abs/1412.6119}{{\ttfamily 1412.6119}}].

\bibitem{Bramante:2017ulk}
J.~Bramante, T.~Linden and Y.-D. Tsai, \emph{{Searching for dark matter with
  neutron star mergers and quiet kilonovae}},
  \href{https://doi.org/10.1103/PhysRevD.97.055016}{\emph{Phys. Rev. D}
  {\bfseries 97} (2018) 055016}
  [\href{https://arxiv.org/abs/1706.00001}{{\ttfamily 1706.00001}}].

\bibitem{Takhistov:2020vxs}
V.~Takhistov, G.~M. Fuller and A.~Kusenko, \emph{{Test for the Origin of Solar
  Mass Black Holes}},
  \href{https://doi.org/10.1103/PhysRevLett.126.071101}{\emph{Phys. Rev. Lett.}
  {\bfseries 126} (2021) 071101}
  [\href{https://arxiv.org/abs/2008.12780}{{\ttfamily 2008.12780}}].

\bibitem{Garani:2021gvc}
R.~Garani, D.~Levkov and P.~Tinyakov, \emph{{Solar mass black holes from
  neutron stars and bosonic dark matter}},
  \href{https://doi.org/10.1103/PhysRevD.105.063019}{\emph{Phys. Rev. D}
  {\bfseries 105} (2022) 063019}
  [\href{https://arxiv.org/abs/2112.09716}{{\ttfamily 2112.09716}}].

\bibitem{Garani:2022quc}
R.~Garani, M.~H.~G. Tytgat and J.~Vandecasteele, \emph{{Condensed dark matter
  with a Yukawa interaction}},
  \href{https://doi.org/10.1103/PhysRevD.106.116003}{\emph{Phys. Rev. D}
  {\bfseries 106} (2022) 116003}
  [\href{https://arxiv.org/abs/2207.06928}{{\ttfamily 2207.06928}}].

\bibitem{Steigerwald:2022pjo}
H.~Steigerwald, V.~Marra and S.~Profumo, \emph{{Revisiting constraints on
  asymmetric dark matter from collapse in white dwarf stars}},
  \href{https://doi.org/10.1103/PhysRevD.105.083507}{\emph{Phys. Rev. D}
  {\bfseries 105} (2022) 083507}
  [\href{https://arxiv.org/abs/2203.09054}{{\ttfamily 2203.09054}}].

\bibitem{Dexter:2013xga}
J.~Dexter and R.~M. O'Leary, \emph{{The Peculiar Pulsar Population of the
  Central Parsec}},
  \href{https://doi.org/10.1088/2041-8205/783/1/L7}{\emph{Astrophys. J. Lett.}
  {\bfseries 783} (2014) L7} [\href{https://arxiv.org/abs/1310.7022}{{\ttfamily
  1310.7022}}].

\bibitem{Takhistov:2017bpt}
V.~Takhistov, \emph{{Transmuted Gravity Wave Signals from Primordial Black
  Holes}}, \href{https://doi.org/10.1016/j.physletb.2018.05.026}{\emph{Phys.
  Lett. B} {\bfseries 782} (2018) 77}
  [\href{https://arxiv.org/abs/1707.05849}{{\ttfamily 1707.05849}}].

\bibitem{Bramante:2017xlb}
J.~Bramante, A.~Delgado and A.~Martin, \emph{{Multiscatter stellar capture of
  dark matter}}, \href{https://doi.org/10.1103/PhysRevD.96.063002}{\emph{Phys.
  Rev. D} {\bfseries 96} (2017) 063002}
  [\href{https://arxiv.org/abs/1703.04043}{{\ttfamily 1703.04043}}].

\bibitem{Dasgupta:2019juq}
B.~Dasgupta, A.~Gupta and A.~Ray, \emph{{Dark matter capture in celestial
  objects: Improved treatment of multiple scattering and updated constraints
  from white dwarfs}},
  \href{https://doi.org/10.1088/1475-7516/2019/08/018}{\emph{JCAP} {\bfseries
  08} (2019) 018} [\href{https://arxiv.org/abs/1906.04204}{{\ttfamily
  1906.04204}}].

\bibitem{Anzuini:2021lnv}
F.~Anzuini, N.~F. Bell, G.~Busoni, T.~F. Motta, S.~Robles, A.~W. Thomas et~al.,
  \emph{{Improved treatment of dark matter capture in neutron stars III:
  nucleon and exotic targets}},
  \href{https://doi.org/10.1088/1475-7516/2021/11/056}{\emph{JCAP} {\bfseries
  11} (2021) 056} [\href{https://arxiv.org/abs/2108.02525}{{\ttfamily
  2108.02525}}].

\bibitem{Bell:2020obw}
N.~F. Bell, G.~Busoni, T.~F. Motta, S.~Robles, A.~W. Thomas and M.~Virgato,
  \emph{{Nucleon Structure and Strong Interactions in Dark Matter Capture in
  Neutron Stars}},
  \href{https://doi.org/10.1103/PhysRevLett.127.111803}{\emph{Phys. Rev. Lett.}
  {\bfseries 127} (2021) 111803}
  [\href{https://arxiv.org/abs/2012.08918}{{\ttfamily 2012.08918}}].

\bibitem{East:2019dxt}
W.~E. East and L.~Lehner, \emph{{Fate of a neutron star with an endoparasitic
  black hole and implications for dark matter}},
  \href{https://doi.org/10.1103/PhysRevD.100.124026}{\emph{Phys. Rev. D}
  {\bfseries 100} (2019) 124026}
  [\href{https://arxiv.org/abs/1909.07968}{{\ttfamily 1909.07968}}].

\bibitem{Baumgarte:2021thx}
T.~W. Baumgarte and S.~L. Shapiro, \emph{{Neutron Stars Harboring a Primordial
  Black Hole: Maximum Survival Time}},
  \href{https://doi.org/10.1103/PhysRevD.103.L081303}{\emph{Phys. Rev. D}
  {\bfseries 103} (2021) L081303}
  [\href{https://arxiv.org/abs/2101.12220}{{\ttfamily 2101.12220}}].

\bibitem{Richards:2021upu}
C.~B. Richards, T.~W. Baumgarte and S.~L. Shapiro, \emph{{Accretion onto a
  small black hole at the center of a neutron star}},
  \href{https://doi.org/10.1103/PhysRevD.103.104009}{\emph{Phys. Rev. D}
  {\bfseries 103} (2021) 104009}
  [\href{https://arxiv.org/abs/2102.09574}{{\ttfamily 2102.09574}}].

\bibitem{Schnauck:2021hlm}
S.~C. Schnauck, T.~W. Baumgarte and S.~L. Shapiro, \emph{{Accretion onto black
  holes inside neutron stars with piecewise-polytropic equations of state:
  Analytic and numerical treatments}},
  \href{https://doi.org/10.1103/PhysRevD.104.123021}{\emph{Phys. Rev. D}
  {\bfseries 104} (2021) 123021}
  [\href{https://arxiv.org/abs/2110.08285}{{\ttfamily 2110.08285}}].

\bibitem{Giffin:2021kgb}
P.~Giffin, J.~Lloyd, S.~D. McDermott and S.~Profumo, \emph{{Neutron Star
  Quantum Death by Small Black Holes}},
  \href{https://arxiv.org/abs/2105.06504}{{\ttfamily 2105.06504}}.

\bibitem{Navarro:1995iw}
J.~F. Navarro, C.~S. Frenk and S.~D.~M. White, \emph{{The Structure of cold
  dark matter halos}}, \href{https://doi.org/10.1086/177173}{\emph{Astrophys.
  J.} {\bfseries 462} (1996) 563}
  [\href{https://arxiv.org/abs/astro-ph/9508025}{{\ttfamily
  astro-ph/9508025}}].

\bibitem{Navarro:1996gj}
J.~F. Navarro, C.~S. Frenk and S.~D.~M. White, \emph{{A Universal density
  profile from hierarchical clustering}},
  \href{https://doi.org/10.1086/304888}{\emph{Astrophys. J.} {\bfseries 490}
  (1997) 493} [\href{https://arxiv.org/abs/astro-ph/9611107}{{\ttfamily
  astro-ph/9611107}}].

\bibitem{Taylor:2012db}
S.~R. Taylor and J.~R. Gair, \emph{{Cosmology with the lights off: standard
  sirens in the Einstein Telescope era}},
  \href{https://doi.org/10.1103/PhysRevD.86.023502}{\emph{Phys. Rev. D}
  {\bfseries 86} (2012) 023502}
  [\href{https://arxiv.org/abs/1204.6739}{{\ttfamily 1204.6739}}].

\bibitem{Madau:2014bja}
P.~Madau and M.~Dickinson, \emph{{Cosmic Star Formation History}},
  \href{https://doi.org/10.1146/annurev-astro-081811-125615}{\emph{Ann. Rev.
  Astron. Astrophys.} {\bfseries 52} (2014) 415}
  [\href{https://arxiv.org/abs/1403.0007}{{\ttfamily 1403.0007}}].

\bibitem{OShaughnessy:2007brt}
R.~W. O'Shaughnessy, V.~Kalogera and K.~Belczynski, \emph{{Short Gamma-Ray
  Bursts and Binary Mergers in Spiral and Elliptical Galaxies: Redshift
  Distribution and Hosts}},
  \href{https://doi.org/10.1086/526334}{\emph{Astrophys. J.} {\bfseries 675}
  (2008) 566} [\href{https://arxiv.org/abs/0706.4139}{{\ttfamily 0706.4139}}].

\bibitem{OShaughnessy:2009szr}
R.~O'Shaughnessy, V.~Kalogera and K.~Belczynski, \emph{{Binary Compact Object
  Coalescence Rates: The Role of Elliptical Galaxies}},
  \href{https://doi.org/10.1088/0004-637X/716/1/615}{\emph{Astrophys. J.}
  {\bfseries 716} (2010) 615}
  [\href{https://arxiv.org/abs/0908.3635}{{\ttfamily 0908.3635}}].

\bibitem{LIGOScientific:2021psn}
{\scshape LIGO Scientific, VIRGO, KAGRA} collaboration, R.~Abbott et~al.,
  \emph{{The population of merging compact binaries inferred using
  gravitational waves through GWTC-3}},
  \href{https://arxiv.org/abs/2111.03634}{{\ttfamily 2111.03634}}.

\bibitem{Acevedo:2020gro}
J.~F. Acevedo, J.~Bramante, A.~Goodman, J.~Kopp and T.~Opferkuch, \emph{{Dark
  Matter, Destroyer of Worlds: Neutrino, Thermal, and Existential Signatures
  from Black Holes in the Sun and Earth}},
  \href{https://doi.org/10.1088/1475-7516/2021/04/026}{\emph{JCAP} {\bfseries
  04} (2021) 026} [\href{https://arxiv.org/abs/2012.09176}{{\ttfamily
  2012.09176}}].

\bibitem{Ray:2023auh}
A.~Ray, \emph{{Celestial Objects as Strongly-Interacting Asymmetric Dark Matter
  Detectors}},  \href{https://arxiv.org/abs/2301.03625}{{\ttfamily
  2301.03625}}.

\bibitem{Boudaud:2014qra}
M.~Boudaud, M.~Cirelli, G.~Giesen and P.~Salati, \emph{{A fussy revisitation of
  antiprotons as a tool for Dark Matter searches}},
  \href{https://doi.org/10.1088/1475-7516/2015/05/013}{\emph{JCAP} {\bfseries
  05} (2015) 013} [\href{https://arxiv.org/abs/1412.5696}{{\ttfamily
  1412.5696}}].

\bibitem{1980ApJS...44...73B}
J.~N. {Bahcall} and R.~M. {Soneira}, \emph{{The universe at faint magnitudes.
  I. Models for the Galaxy and the predicted star counts.}},
  \href{https://doi.org/10.1086/190685}{\emph{The Astrophysical Journal
  Supplement Series} {\bfseries 44} (1980) 73}.

\bibitem{Steidel:1998ja}
C.~C. Steidel, K.~L. Adelberger, M.~Giavalisco, M.~Dickinson and M.~Pettini,
  \emph{{Lyman break galaxies at z $>$ 4 and the evolution of the UV luminosity
  density at high redshift}},
  \href{https://doi.org/10.1086/307363}{\emph{Astrophys. J.} {\bfseries 519}
  (1999) 1} [\href{https://arxiv.org/abs/astro-ph/9811399}{{\ttfamily
  astro-ph/9811399}}].

\bibitem{Porciani:2000ag}
C.~Porciani and P.~Madau, \emph{{On the Association of gamma-ray bursts with
  massive stars: implications for number counts and lensing statistics}},
  \href{https://doi.org/10.1086/319027}{\emph{Astrophys. J.} {\bfseries 548}
  (2001) 522} [\href{https://arxiv.org/abs/astro-ph/0008294}{{\ttfamily
  astro-ph/0008294}}].

\bibitem{2010MNRAS.402..371B}
S.~{Banerjee}, H.~{Baumgardt} and P.~{Kroupa}, \emph{{Stellar-mass black holes
  in star clusters: implications for gravitational wave radiation}},
  \href{https://doi.org/10.1111/j.1365-2966.2009.15880.x}{\emph{Monthly Notices
  of the Royal Astronomical Society} {\bfseries 402} (2010) 371}
  [\href{https://arxiv.org/abs/0910.3954}{{\ttfamily 0910.3954}}].

\bibitem{Dominik:2012kk}
M.~Dominik, K.~Belczynski, C.~Fryer, D.~Holz, E.~Berti, T.~Bulik et~al.,
  \emph{{Double Compact Objects I: The Significance of the Common Envelope on
  Merger Rates}},
  \href{https://doi.org/10.1088/0004-637X/759/1/52}{\emph{Astrophys. J.}
  {\bfseries 759} (2012) 52} [\href{https://arxiv.org/abs/1202.4901}{{\ttfamily
  1202.4901}}].

\bibitem{Vitale:2018yhm}
S.~Vitale, W.~M. Farr, K.~Ng and C.~L. Rodriguez, \emph{{Measuring the star
  formation rate with gravitational waves from binary black holes}},
  \href{https://doi.org/10.3847/2041-8213/ab50c0}{\emph{Astrophys. J. Lett.}
  {\bfseries 886} (2019) L1}
  [\href{https://arxiv.org/abs/1808.00901}{{\ttfamily 1808.00901}}].

\bibitem{Santoliquido:2020axb}
F.~Santoliquido, M.~Mapelli, N.~Giacobbo, Y.~Bouffanais and M.~C. Artale,
  \emph{{The cosmic merger rate density of compact objects: impact of star
  formation, metallicity, initial mass function and binary evolution}},
  \href{https://doi.org/10.1093/mnras/stab280}{\emph{Mon. Not. Roy. Astron.
  Soc.} {\bfseries 502} (2021) 4877}
  [\href{https://arxiv.org/abs/2009.03911}{{\ttfamily 2009.03911}}].

\bibitem{Mukherjee:2021qam}
S.~Mukherjee, T.~Broadhurst, J.~M. Diego, J.~Silk and G.~F. Smoot,
  \emph{{Impact of astrophysical binary coalescence time-scales on the rate of
  lensed gravitational wave events}},
  \href{https://doi.org/10.1093/mnras/stab1980}{\emph{Mon. Not. Roy. Astron.
  Soc.} {\bfseries 506} (2021) 3751}
  [\href{https://arxiv.org/abs/2106.00392}{{\ttfamily 2106.00392}}].

\bibitem{Steigman:2012nb}
G.~Steigman, B.~Dasgupta and J.~F. Beacom, \emph{{Precise Relic WIMP Abundance
  and its Impact on Searches for Dark Matter Annihilation}},
  \href{https://doi.org/10.1103/PhysRevD.86.023506}{\emph{Phys. Rev. D}
  {\bfseries 86} (2012) 023506}
  [\href{https://arxiv.org/abs/1204.3622}{{\ttfamily 1204.3622}}].

\bibitem{Barsanti:2021ydd}
S.~Barsanti, V.~De~Luca, A.~Maselli and P.~Pani, \emph{{Detecting Subsolar-Mass
  Primordial Black Holes in Extreme Mass-Ratio Inspirals with LISA and Einstein
  Telescope}},
  \href{https://doi.org/10.1103/PhysRevLett.128.111104}{\emph{Phys. Rev. Lett.}
  {\bfseries 128} (2022) 111104}
  [\href{https://arxiv.org/abs/2109.02170}{{\ttfamily 2109.02170}}].

\bibitem{Joglekar:2019vzy}
A.~Joglekar, N.~Raj, P.~Tanedo and H.-B. Yu, \emph{{Relativistic capture of
  dark matter by electrons in neutron stars}},
  \href{https://doi.org/10.1016/j.physletb.2020.135767}{\emph{Phys. Lett.}
  {\bfseries B} (2020) 135767}
  [\href{https://arxiv.org/abs/1911.13293}{{\ttfamily 1911.13293}}].

\bibitem{Bell:2020lmm}
N.~F. Bell, G.~Busoni, S.~Robles and M.~Virgato, \emph{{Improved Treatment of
  Dark Matter Capture in Neutron Stars II: Leptonic Targets}},
  \href{https://doi.org/10.1088/1475-7516/2021/03/086}{\emph{JCAP} {\bfseries
  03} (2021) 086} [\href{https://arxiv.org/abs/2010.13257}{{\ttfamily
  2010.13257}}].

\bibitem{Nunes:2021jsk}
R.~C. Nunes, \emph{{Search for Sub-Solar Mass Binaries with Einstein Telescope
  and Cosmic Explorer}},
  \href{https://doi.org/10.3390/e24020262}{\emph{Entropy} {\bfseries 24} (2022)
  262} [\href{https://arxiv.org/abs/2109.05910}{{\ttfamily 2109.05910}}].

\bibitem{LIGOScientific:2022hai}
{\scshape LIGO Scientific, VIRGO, KAGRA} collaboration, R.~Abbott et~al.,
  \emph{{Search for subsolar-mass black hole binaries in the second part of
  Advanced LIGO's and Advanced Virgo's third observing run}},
  \href{https://arxiv.org/abs/2212.01477}{{\ttfamily 2212.01477}}.

\bibitem{LZ:2022lsv}
{\scshape LZ} collaboration, J.~Aalbers et~al., \emph{{First Dark Matter Search
  Results from the LUX-ZEPLIN (LZ) Experiment}},
  \href{https://doi.org/10.1103/PhysRevLett.131.041002}{\emph{Phys. Rev. Lett.}
  {\bfseries 131} (2023) 041002}
  [\href{https://arxiv.org/abs/2207.03764}{{\ttfamily 2207.03764}}].

\bibitem{OHare:2021utq}
C.~A.~J. O'Hare, \emph{{New Definition of the Neutrino Floor for Direct Dark
  Matter Searches}},
  \href{https://doi.org/10.1103/PhysRevLett.127.251802}{\emph{Phys. Rev. Lett.}
  {\bfseries 127} (2021) 251802}
  [\href{https://arxiv.org/abs/2109.03116}{{\ttfamily 2109.03116}}].

\bibitem{Ozel:2016oaf}
F.~\"Ozel and P.~Freire, \emph{{Masses, Radii, and the Equation of State of
  Neutron Stars}},
  \href{https://doi.org/10.1146/annurev-astro-081915-023322}{\emph{Ann. Rev.
  Astron. Astrophys.} {\bfseries 54} (2016) 401}
  [\href{https://arxiv.org/abs/1603.02698}{{\ttfamily 1603.02698}}].

\bibitem{2013PASP..125..306F}
D.~{Foreman-Mackey}, D.~W. {Hogg}, D.~{Lang} and J.~{Goodman}, \emph{{emcee:
  The MCMC Hammer}}, \href{https://doi.org/10.1086/670067}{\emph{Publications
  of the Astronomical Society of the Pacific} {\bfseries 125} (2013) 306}
  [\href{https://arxiv.org/abs/1202.3665}{{\ttfamily 1202.3665}}].

\bibitem{LVKObs}
\url{https://observing.docs.ligo.org/plan/}.

\bibitem{Maggiore:2019uih}
M.~Maggiore et~al., \emph{{Science Case for the Einstein Telescope}},
  \href{https://doi.org/10.1088/1475-7516/2020/03/050}{\emph{JCAP} {\bfseries
  03} (2020) 050} [\href{https://arxiv.org/abs/1912.02622}{{\ttfamily
  1912.02622}}].

\bibitem{Evans:2021gyd}
M.~Evans et~al., \emph{{A Horizon Study for Cosmic Explorer: Science,
  Observatories, and Community}},
  \href{https://arxiv.org/abs/2109.09882}{{\ttfamily 2109.09882}}.

\bibitem{LIGOScientific:2005fbz}
{\scshape LIGO Scientific} collaboration, B.~Abbott et~al., \emph{{Search for
  gravitational waves from primordial black hole binary coalescences in the
  galactic halo}},
  \href{https://doi.org/10.1103/PhysRevD.72.082002}{\emph{Phys. Rev. D}
  {\bfseries 72} (2005) 082002}
  [\href{https://arxiv.org/abs/gr-qc/0505042}{{\ttfamily gr-qc/0505042}}].

\bibitem{LIGOScientific:2007npa}
{\scshape LIGO Scientific} collaboration, B.~Abbott et~al., \emph{{Search for
  gravitational waves from binary inspirals in S3 and S4 LIGO data}},
  \href{https://doi.org/10.1103/PhysRevD.77.062002}{\emph{Phys. Rev. D}
  {\bfseries 77} (2008) 062002}
  [\href{https://arxiv.org/abs/0704.3368}{{\ttfamily 0704.3368}}].

\bibitem{LIGOScientific:2018glc}
{\scshape LIGO Scientific, Virgo} collaboration, B.~P. Abbott et~al.,
  \emph{{Search for Subsolar-Mass Ultracompact Binaries in Advanced
  LIGO\textquoteright{}s First Observing Run}},
  \href{https://doi.org/10.1103/PhysRevLett.121.231103}{\emph{Phys. Rev. Lett.}
  {\bfseries 121} (2018) 231103}
  [\href{https://arxiv.org/abs/1808.04771}{{\ttfamily 1808.04771}}].

\bibitem{LIGOScientific:2019kan}
{\scshape LIGO Scientific, Virgo} collaboration, B.~P. Abbott et~al.,
  \emph{{Search for Subsolar Mass Ultracompact Binaries in Advanced
  LIGO\textquoteright{}s Second Observing Run}},
  \href{https://doi.org/10.1103/PhysRevLett.123.161102}{\emph{Phys. Rev. Lett.}
  {\bfseries 123} (2019) 161102}
  [\href{https://arxiv.org/abs/1904.08976}{{\ttfamily 1904.08976}}].

\bibitem{LIGOScientific:2021job}
{\scshape LIGO Scientific, VIRGO, KAGRA} collaboration, R.~Abbott et~al.,
  \emph{{Search for Subsolar-Mass Binaries in the First Half of Advanced
  LIGO\textquoteright{}s and Advanced Virgo\textquoteright{}s Third Observing
  Run}}, \href{https://doi.org/10.1103/PhysRevLett.129.061104}{\emph{Phys. Rev.
  Lett.} {\bfseries 129} (2022) 061104}
  [\href{https://arxiv.org/abs/2109.12197}{{\ttfamily 2109.12197}}].

\bibitem{Nitz:2021mzz}
A.~H. Nitz and Y.-F. Wang, \emph{{Search for gravitational waves from the
  coalescence of sub-solar mass and eccentric compact binaries}},
  \href{https://arxiv.org/abs/2102.00868}{{\ttfamily 2102.00868}}.

\bibitem{Nitz:2021vqh}
A.~H. Nitz and Y.-F. Wang, \emph{{Search for Gravitational Waves from the
  Coalescence of Subsolar-Mass Binaries in the First Half of Advanced LIGO and
  Virgo\textquoteright{}s Third Observing Run}},
  \href{https://doi.org/10.1103/PhysRevLett.127.151101}{\emph{Phys. Rev. Lett.}
  {\bfseries 127} (2021) 151101}
  [\href{https://arxiv.org/abs/2106.08979}{{\ttfamily 2106.08979}}].

\bibitem{Nitz:2022ltl}
A.~H. Nitz and Y.-F. Wang, \emph{{Broad search for gravitational waves from
  subsolar-mass binaries through LIGO and Virgo\textquoteright{}s third
  observing run}},
  \href{https://doi.org/10.1103/PhysRevD.106.023024}{\emph{Phys. Rev. D}
  {\bfseries 106} (2022) 023024}
  [\href{https://arxiv.org/abs/2202.11024}{{\ttfamily 2202.11024}}].

\bibitem{Phukon:2021cus}
K.~S. Phukon, G.~Baltus, S.~Caudill, S.~Clesse, A.~Depasse, M.~Fays et~al.,
  \emph{{The hunt for sub-solar primordial black holes in low mass ratio
  binaries is open}},  \href{https://arxiv.org/abs/2105.11449}{{\ttfamily
  2105.11449}}.

\bibitem{Singh:2020wiq}
D.~Singh, M.~Ryan, R.~Magee, T.~Akhter, S.~Shandera, D.~Jeong et~al.,
  \emph{{Gravitational-wave limit on the Chandrasekhar mass of dark matter}},
  \href{https://doi.org/10.1103/PhysRevD.104.044015}{\emph{Phys. Rev. D}
  {\bfseries 104} (2021) 044015}
  [\href{https://arxiv.org/abs/2009.05209}{{\ttfamily 2009.05209}}].

\bibitem{Singh:2022wvw}
D.~Singh, A.~Gupta, E.~Berti, S.~Reddy and B.~S. Sathyaprakash,
  \emph{{Constraining properties of asymmetric dark matter candidates from
  gravitational-wave observations}},
  \href{https://arxiv.org/abs/2210.15739}{{\ttfamily 2210.15739}}.

\end{thebibliography}\endgroup

\clearpage
\newpage
\maketitle
\onecolumngrid
\begin{center}
	\textbf{\Large Supplemental Material}
	 \bigskip\\
		\textbf{\large Can LIGO Detect Non-Annihilating Dark Matter?}
		 \medskip\\
   {Sulagna Bhattacharya, Basudeb Dasgupta, Ranjan Laha, and Anupam Ray}
\end{center}

\renewcommand{\thesection}{S\arabic{section}}
\renewcommand{\theequation}{S\arabic{equation}}
\renewcommand{\thefigure}{S\arabic{figure}}
\renewcommand{\thetable}{S\arabic{table}}
\renewcommand{\thepage}{S\arabic{page}}
\setcounter{equation}{0}
\setcounter{figure}{0}
\setcounter{table}{0}
\setcounter{page}{1}
\setcounter{secnumdepth}{4}

In the Supplemental Material, we briefly review dark matter (DM)-induced neutron star (NS) collapse conditions, and show how the transmuted black hole merger rate depends on various model assumptions, e.g., progenitor properties (such as mass, radius, and core temperature of the NSs), astrophysical uncertainties (such as cosmic star formation rate, DM density profile in the halos), and DM properties.
\section{DM Accretion induced NS Collapse}

\begin{figure}[!h]
	\centering
	\includegraphics[width=0.85 \columnwidth]{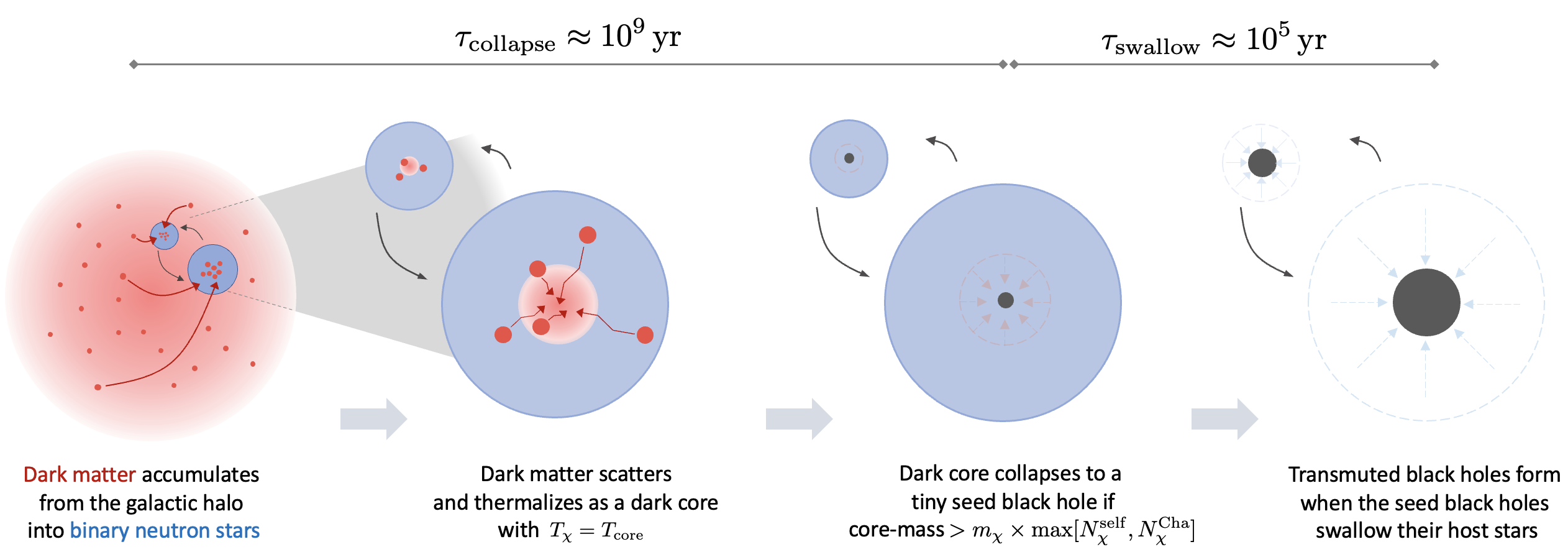}
	\caption{Schematic diagram for transmutation of a binary neutron star system via accumulation of non-annihilating DM. The transmutation timescales are estimated for (bosonic) DM mass of $m_{\chi}=10^5\rm\, GeV$ and DM-nucleon scattering cross-section of $\sigma_{\chi n}=10^{-45}\rm\,cm^2$ with an ambient DM density of 0.4 GeV/cm$^3$. $\tau_{\rm{collapse}}$ denotes the timescale for  seed BH formation and $\tau_{\rm{swallow}}$ denotes the timescale by which the seed BH destroys the star.  For the progenitors, we take mass $(M_{\rm{NS}}) = 1.35\,M_{\odot}$, radius $(R_{\rm{NS}}) = 10\,{\rm{km}}$, and core temperature $(T_{\rm core})=2.1\times 10^6\,\rm{K}$.}
	\label{fig: fig}
\end{figure}
Non-annihilating DM particles, owing to their interactions with the stellar nucleons, gradually  accumulate inside NSs. For small DM-nucleon interactions (optically thin regime), the accumulation rate is~\cite{McDermott:2011jp,Garani:2018kkd} 
\begin{align}
	C = 1.4\times 10^{20}\,{\rm s}^{-1}\,\Big(\tfrac{\rho_{\chi}}{0.4\, \rm{GeV\,cm^{-3}}}\Big) \Big(\tfrac{10^5\, \rm{GeV}}{m_{\chi}}\Big)\Big(\tfrac{\sigma_{\chi n}}{10^{-45}\,\rm{cm^2}}\Big)
	\left(1-\tfrac{1-e^{-A^2}}{A^2}\right)\,\left(\tfrac{v_{\rm{esc}}}{1.9\times10^{5}\,\rm km\,s^{-1}}\right)^2\left(\tfrac{220\,\rm km\,s^{-1}}{\bar{v}_{\rm gal}}\right)^2\,,
\end{align}
where  $\rho_{\chi}$ denotes the ambient DM density,  $m_\chi$ denotes the DM mass, and $\sigma_{\chi n}$ denotes the total interaction cross-section with nucleons. The factor involving $A^2={6\,m_{\chi}m_n}{v^2_{\rm{esc}}}/{\bar{v}^2_{\rm gal}}{(m_{\chi}-m_n)^2}$ accounts for inefficient momentum transfers in the collisions, given NS escape speed $v_{\rm esc}$ and typical dark matter speed ${\bar{v}_{\rm gal}}$ in the galactic halo. For heavy DM masses ($m_{\chi} \gg 10^7$ GeV), it causes a kinematic suppression of $A^2/2$, implying $C \sim 1/m^2_{\chi}$ for $m_{\chi} \gg 10^7$\,GeV.

After being captured by the star, the DM particles sink towards the center of the NS and form an approximately isothermal dark core in a relatively short timescale~\cite{McDermott:2011jp,Kouvaris:2011fi,Garani:2018kkd,Dasgupta:2020dik}, e.g., of order $10^{-3}\,{\rm yr}$ for bosonic DM with $m_{\chi}=10^5\rm\, GeV$ and $\sigma_{\chi n}=10^{-45}\rm\,cm^2$. The radius of the thermalized dark core is $r_{\rm th}=\left(\tfrac{9k_{\rm B}T_{\rm core_{}}}{4\pi G_{\rm N}\rho_{\rm NS}m_{\chi}}\right)^{1/2}$ as dictated by the virial theorem, and can be very small for heavier DM masses. Quantitatively, for typical NS parameters, it can be as tiny as $\sim 5$\,cm for $m_{\chi} = 10^5$\,GeV, as it reduces as $1/m^{1/2}_{\chi}$, resulting in an enormous core-density. Once the core-density exceeds the critical density for BH formation, the dark core undergoes gravitational collapse and forms a tiny BH inside the stellar core (see also Refs.~\cite{Acevedo:2020gro,Ray:2023auh} for BH formation inside non-compact stellar objects via accumulation of strongly-interacting heavy DM). The timescale for such gravitational collapse  is~\cite{McDermott:2011jp,Kouvaris:2011fi,Garani:2018kkd,Dasgupta:2020dik}
\begin{align}
	{\tau_{\rm collapse}} = C^{-1}N^{\rm{BH}}_{\chi}\,,
\end{align}
where
\begin{align}
	N^{\rm{BH}}_{\chi}= \max \left[ N^{\rm{self}}_{\chi}, N^{\rm{Cha}}_{\chi}\right]\,,
\end{align}
is the number of DM particles that need to be captured and thermalized to create a BH. The first term, $N^{\rm{self}}_{\chi}$, denotes the numbers required for initiating self-gravitating Jeans instability, and is set by the condition that DM density has to exceed the
corresponding baryonic density within the core. It is independent of the spin-statistics of DM particles, and is given by~\cite{McDermott:2011jp,Kouvaris:2011fi,Garani:2018kkd,Dasgupta:2020dik}
\begin{align}
	N^{\rm{self}}_{\chi}=2.1\times 10^{36}\left(\tfrac{10^5\,\mathrm{GeV}}{m_{\chi}}\right)^{5/2}\left(\tfrac{T_{\mathrm{core}}}{2.1 \times 10^6\,\mathrm{K}}\right)^{3/2}\,.
\end{align}
The second term, $N^{\rm{Cha}}_{\chi}$, denotes the Chandrasekhar limit, and represents the threshold after which the self-
gravitating collapse is no longer sustainable against quantum degeneracy pressure. Since the quantum degeneracy pressure for fermionic/bosonic DM behaves differently, Chandrasekhar limit depends on the spin-statistics of DM particles, and is given by~\cite{McDermott:2011jp,Kouvaris:2011fi,Garani:2018kkd,Dasgupta:2020dik}  
\begin{align}
	&N^{\rm{Cha}}_{\chi-\mathrm{fermion}}= 1.8\times 10^{42}\left(\tfrac{10^5\,\mathrm{GeV}}{m_{\chi}}\right)^3\,,{~\rm and~}\\
	&N^{\rm{Cha}}_{\chi-\mathrm{boson}}= 9.5\times 10^{27}\left(\tfrac{10^5\,\mathrm{GeV}}{m_{\chi}}\right)^2\,.
\end{align}
Therefore, the DM accretion induced collapse time-scales for a $1.35\,M_{\odot}$ host NS are,
\begin{align}
	\tau_{\rm collapse}|_{\rm boson}= 4.8\times 10^8 \,{\rm years}\left(\tfrac{T_{\rm core}}{2.1\times 10^6\, {\rm K}}\right)^{3/2}\left(\tfrac{10^5\,{\rm{GeV}}}{m_\chi}\right)^{3/2}\left(\tfrac{0.4 \,{\rm GeV\, cm^{-3}}}{\rho_\chi}\right)\left(\tfrac{10^{-45}\,{\rm cm^2}}{\sigma_{\chi n}}\right)\,,
\end{align}
\begin{align}
	\tau_{\rm collapse}|_{\rm fermion}= 1.9\times 10^{10}\,{\rm years}\left(\tfrac{10^8\,{\rm{GeV}}}{m_\chi}\right)\left(\tfrac{0.4\, {\rm GeV\, cm^{-3}}}{\rho_\chi}\right)\left(\tfrac{10^{-45}\,{\rm cm^2}}{\sigma_{\chi n}}\right)\,.
\end{align}
One finds that $\tau_{\rm collapse}$ is larger for smaller DM mass, and exceeds the age of oldest NSs for $m_\chi\lesssim\mathcal{O} (10^4)$\,GeV (for bosons) and $\mathcal{O} (10^{8})$\,GeV (for fermions), suggesting a minimum DM mass for successful transmutation. In the following, we explore the  transmutation criterion for a possible Bose-Einstein condensate formation inside the progenitors. A schematic diagram for the  transmutation process is depicted in Fig.\,\ref{fig: fig}.
\section{Bosonic DM with BEC formation}
Bosonic DM particles can macroscopically occupy the ground state at  very low temperatures, and form a Bose-Einstein Condensate (BEC). This occurs when the core temperature ($T_{\rm core}$) of the progenitors is lower than the critical temperature for condensation ($T_{\rm crit}$). For $N_{\chi}$ number of captured DM particles within the thermalization volume, the critical temperature is given by~\cite{McDermott:2011jp,Garani:2018kkd}
\begin{equation}
	T_{\rm crit} = \frac{2\pi}{m_{\chi}}\left(\frac{3 N_{\chi}}{4\pi r_{\rm th}^3 \zeta[3/2]}\right)^{2/3}\,,
\end{equation}
where $\zeta[3/2]\approx 2.612$ is the Riemann-Zeta function. It is evident that for sufficiently light DM, a BEC can form more easily because the critical temperature for condensation is higher for lower DM masses. Quantitatively, for $\sigma_{\chi n} = 10^{-45}$ cm$^2$ and for typical neutron star parameters, a DM BEC can form for $m_{\chi}<145$ GeV, assuming that DM self-interactions and possibly other interactions are not significant.

Allowing for BEC formation, the number of DM particles in the condensed ground state is
\begin{align}
	N_{\chi}^0 &= N_{\chi}\left(1-\left(\frac{T_{\rm core}}{T_{\rm crit}}\right)^{3/2}\right) = N_{\chi}-3.1\times 10^{40}\left(\tfrac{T_{\rm core}}{2.1 \times 10^6\,\rm K}\right)^3\,.
	\label{eq:Nbec}
\end{align}
Since these ground-state particles have effectively zero temperature, they distribute within a radius of 
\begin{align}
	r_{\rm BEC} &= \left(\frac{3}{8\pi G m^2_{\chi}\rho_b} \right)^{1/4}= 5.6\times 10^{-7} \mathrm{cm}\,\left(\tfrac{10^5\, \rm{GeV}}{m_{\chi}}\right)^{1/2}\,,
\end{align}
which is much smaller than the usual thermalization radius. As a consequence, the self-gravitating criterion becomes less stringent 
\begin{align}
	N^{\rm{self}}_{\chi} &= \frac{\tfrac{4\pi}{3}\, r_{\rm BEC}^3\,\rho_{b}}{m_{\chi}}=2.6\times 10^{15}\left(\tfrac{10^5\,\mathrm{GeV}}{m_{\chi}}\right)^{5/2}\,,
\end{align}
and the dark core collapse is determined by the Chandrasekhar limit. This implies that, with BEC formation allowed, the dark core collapse criterion is given by~\cite{McDermott:2011jp,Garani:2018kkd}
\begin{align}
	N_{\chi}^0  \geq N^{\rm{Cha}}_{\chi-\mathrm{boson}}\,.
\end{align}
Therefore, the DM accretion induced collapse time scale for low mass bosonic DM particles with BEC formation for a $1.35\,M_{\odot}$ host NS is approximately,
\begin{align}
	\tau_{\rm collapse}|_{\rm BEC}= 1.1\times 10^9 \,{\rm years}\left(\tfrac{10^{-2}\,{\rm{GeV}}}{m_\chi}\right)^{2}\left(\tfrac{0.4 \,{\rm GeV\, cm^{-3}}}{\rho_\chi}\right)\left(\tfrac{10^{-45}\,{\rm cm^2}}{\sigma_{\chi n}}\right)\,.
\end{align}
One finds that $\tau_{\rm collapse}$ is larger for smaller DM mass, and exceeds the age of oldest NSs for $m_\chi\lesssim\mathcal{O} (10^{-3})$\,GeV for bosonic DM particles with allowed BEC formation, suggesting a minimum DM mass for successful transmutation.

\section{Dependence of TBH Merger Rate Density on Model Assumptions}
In this section, we quantify how the TBH merger rate density depends on various model assumptions by varying parameters around the fiducial values noted above. For this section, the $R_{\rm BNS}$ is fixed to $1000\,{\rm Gpc}^{-3}{\rm yr}^{-1}$ for illustration. We restrict the discussion to bosonic DM for brevity; similar results are obtained for fermionic DM. 

\subsection{Progenitor Properties}
\label{sec:NSprops}
\begin{figure*}
	\centering
	\includegraphics[width=0.29\textwidth]{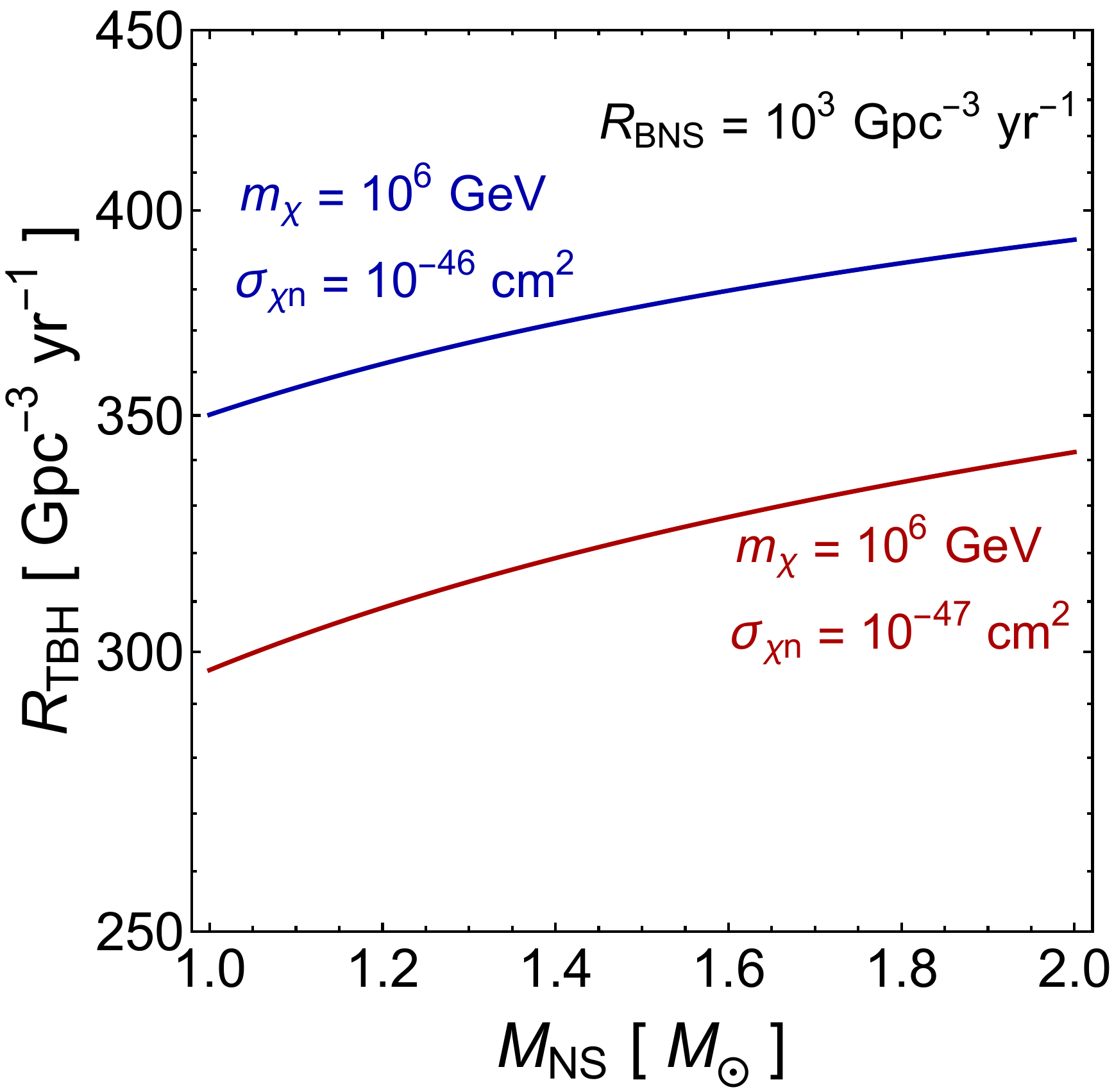}
	\hspace{0.6 cm}
	\includegraphics[width=0.29\textwidth]{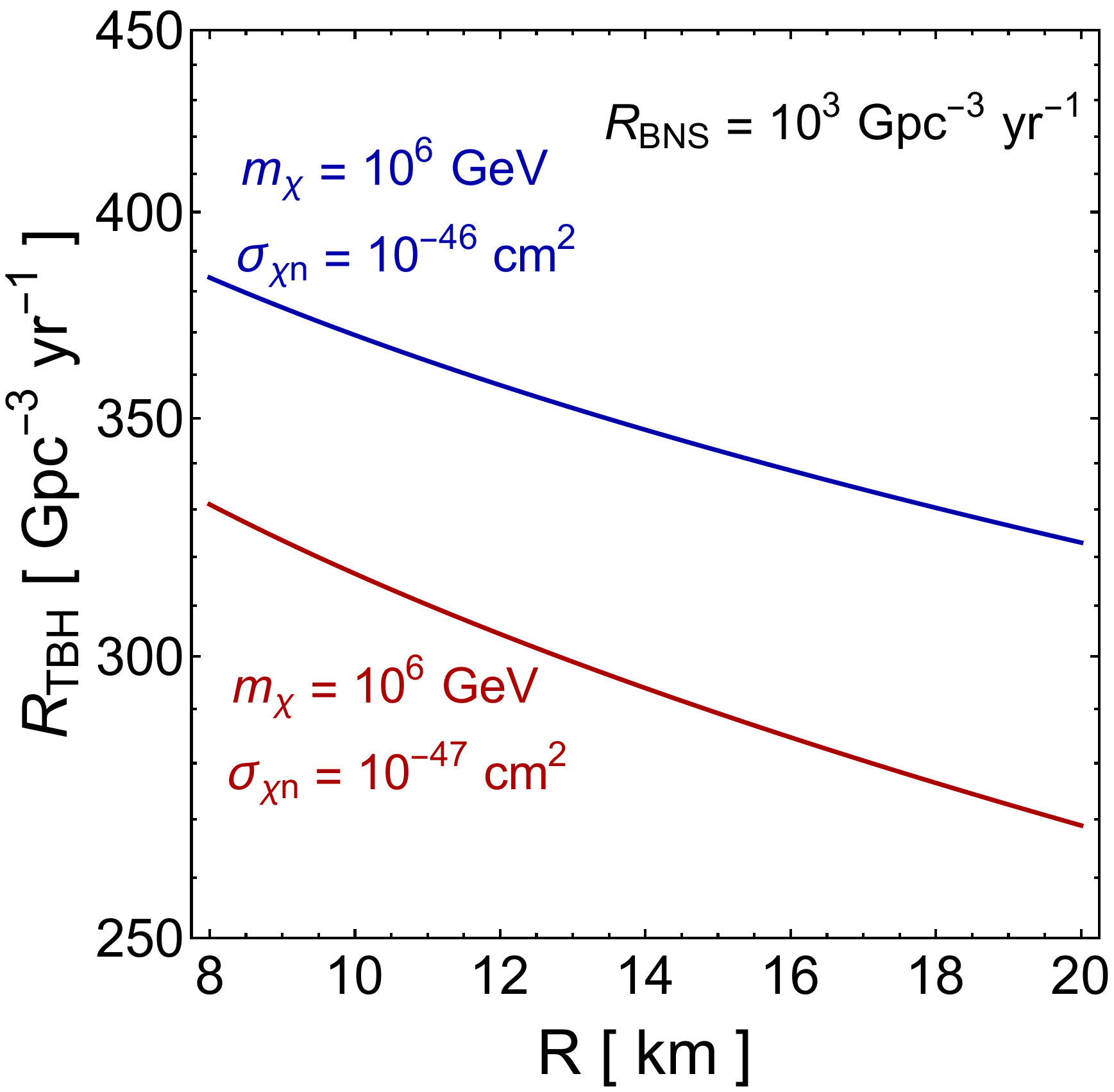}
	\hspace{0.5 cm}
	\includegraphics[width=0.29\textwidth]{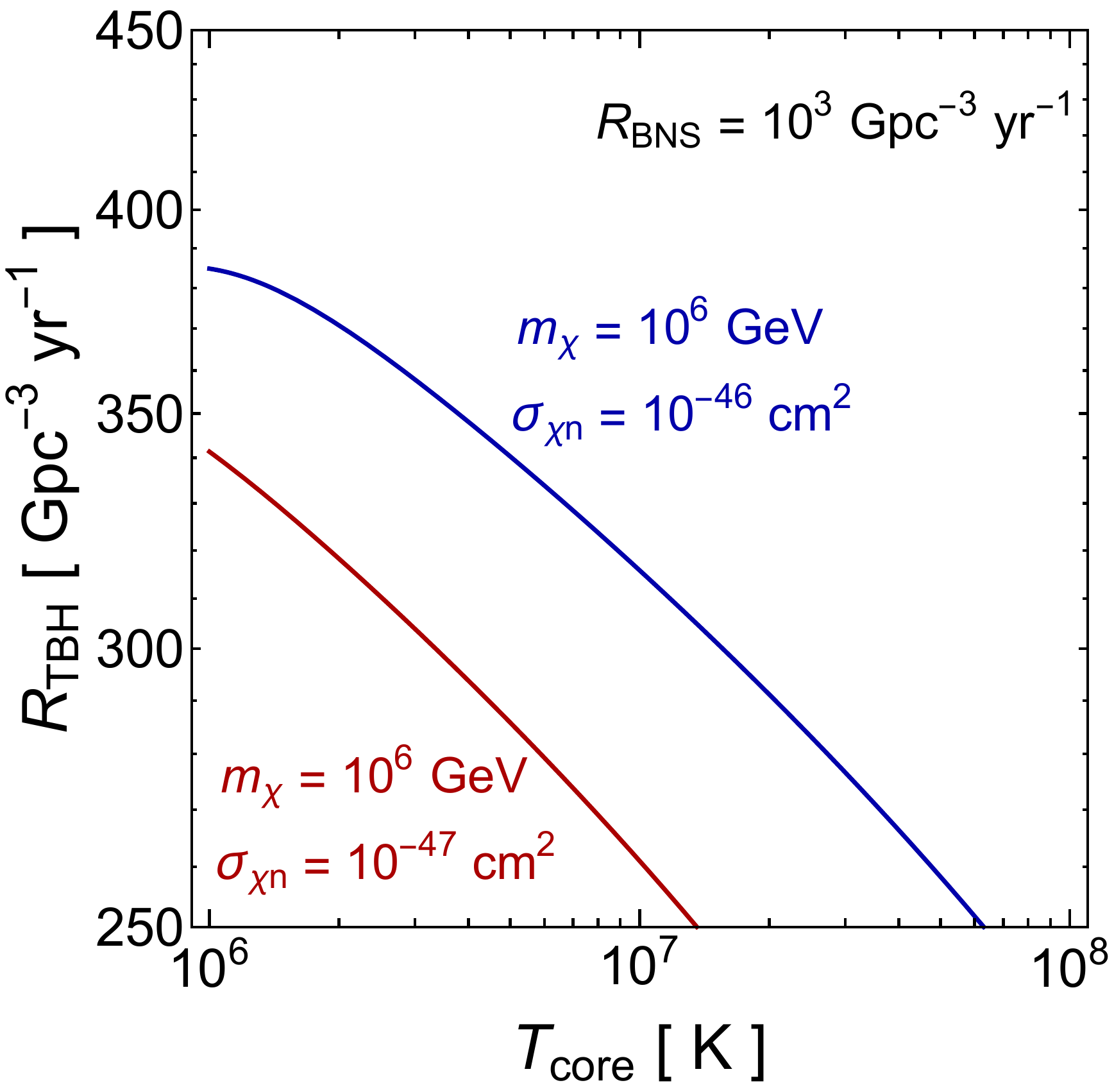}
	\caption[]{Variation of the TBH merger rate density with the properties of the progenitor NS, i.e., mass (left panel), radius (middle panel), and core temperature (right panel). Quantitatively, the NS progenitor properties affects $R_{\rm TBH}$ at a level of $20\%$. See Sec.\,\ref{sec:NSprops} for further discussion.}
	\label{progenitor}
\end{figure*}
\subsubsection{NS Mass}
\label{sec:NSmass}
The capture rate of the incoming DM particles is largely insensitive to the mass of the NSs.  As a consequence, $R_{\rm{TBH}}$ remains almost unaltered with variation in NS mass. Quantitatively, by increasing the NS mass from 1 to 2\,$M_{\odot}$, $R_{\rm{TBH}}$ increases by $\sim$ 12\,(15) \% for $m_{\chi} = 10^6$\,GeV, and $\sigma_{\chi n}=10^{-46} \,(10^{-47})$ cm$^2$. The slight increment of $R_{\rm{TBH}}$ is due to the increased number of stellar target with increase in progenitor mass, implying an increased capture rate.  In Fig.\,\ref{progenitor} (left panel), we show the dependence of the $R_{\rm{TBH}}$ on $M_{\rm NS}$.

\subsubsection{Radius}
\label{sec:NSrad}
The dependence of $R_{\rm TBH}$ on NS radius is solely via $v_{\rm{esc}} [R_{\rm{NS}}]$, which also enters inside $A^2$. Since, the escape velocity scales as 1/${R^{1/2}_{\rm{NS}}}$, the single-scatter capture rate falls off linearly (or quadratically for $m_{\chi} \gg 10^7$ GeV when $A^2\ll 1$), with larger NS radius. In Fig.\,\ref{progenitor} (middle panel), we show the dependence on the NS radius. Quantitatively,  for $m_{\chi} = 10^6$ GeV, and $\sigma_{\chi n}=10^{-46}\,(10^{-47})$ cm$^2$, $R_{\rm{TBH}}$  decreases by 12\,(15)\% with the variation of $R_{\rm{NS}}$ from 10 km to 20 km. In Fig.\,\ref{progenitor} (middle panel) one can see the weak dependence on $R_{\rm NS}$.

\subsubsection{NS Core Temperature}
\label{sec:NStemp}
The core temperature $(T_{\rm{core}})$ of the NS sets the thermalization radius $(r_{\rm{th}})$, within which the captured DM particles are taken to thermalize with the stellar constituents. This radius increases with higher core temperature as $r_{\rm{th}} \sim$ ${T_{\rm{core}}}^{1/2}$. The increment substantially affects the collapse criterion, as the critical number of non-annihilating DM particles in the thermalization volume for ensuring the self-gravitating collapse ($N^{\rm{self}}_{\chi}$) scales cubically with the thermalization radius. With larger core temperature, $r_{\rm{th}}$ increases and results in a much larger $N^{\rm{self}}_{\chi}$, prohibiting BH formation inside the stellar core. It is important to note that, for non-annihilating bosonic DM, $N^{\rm{self}}_{\chi}$ dominates over $N^{\rm{Cha}}_{\chi}$, therefore, the dark collapse is essentially determined by $N^{\rm{self}}_{\chi}$.

In Fig.\,\ref{progenitor} (right panel), we show the core temperature dependence of the TBH merger rate density while keeping the other parameters fixed at their fiducial values. It is evident that TBH merger rate density decreases with larger $T_{\rm{core}}$ as the BH formation gets restrained with higher core temperature. Quantitatively,  for $m_{\chi} = 10^6$ GeV, and $\sigma_{\chi n}=10^{-46}\,(10^{-47})$ cm$^2$, $R_{\rm{TBH}}$  decreases by $\sim$ 18\,(23)\% as $T_{\rm{core}}$ increases from 10$^6$ K to 10$^7$\,K.

To summarize, we conclude that possible variations of the progenitor properties have a mild impact on the TBH merger rate density. Quantitatively, $R_{\rm{TBH}}$ varies by at most 20\% because of progenitor properties.

\subsection{Astrophysical Uncertainties}
\label{sec:astroprop}
\begin{figure*}
	\centering
	\hspace{-0.2cm}\includegraphics[width=0.36\textwidth]{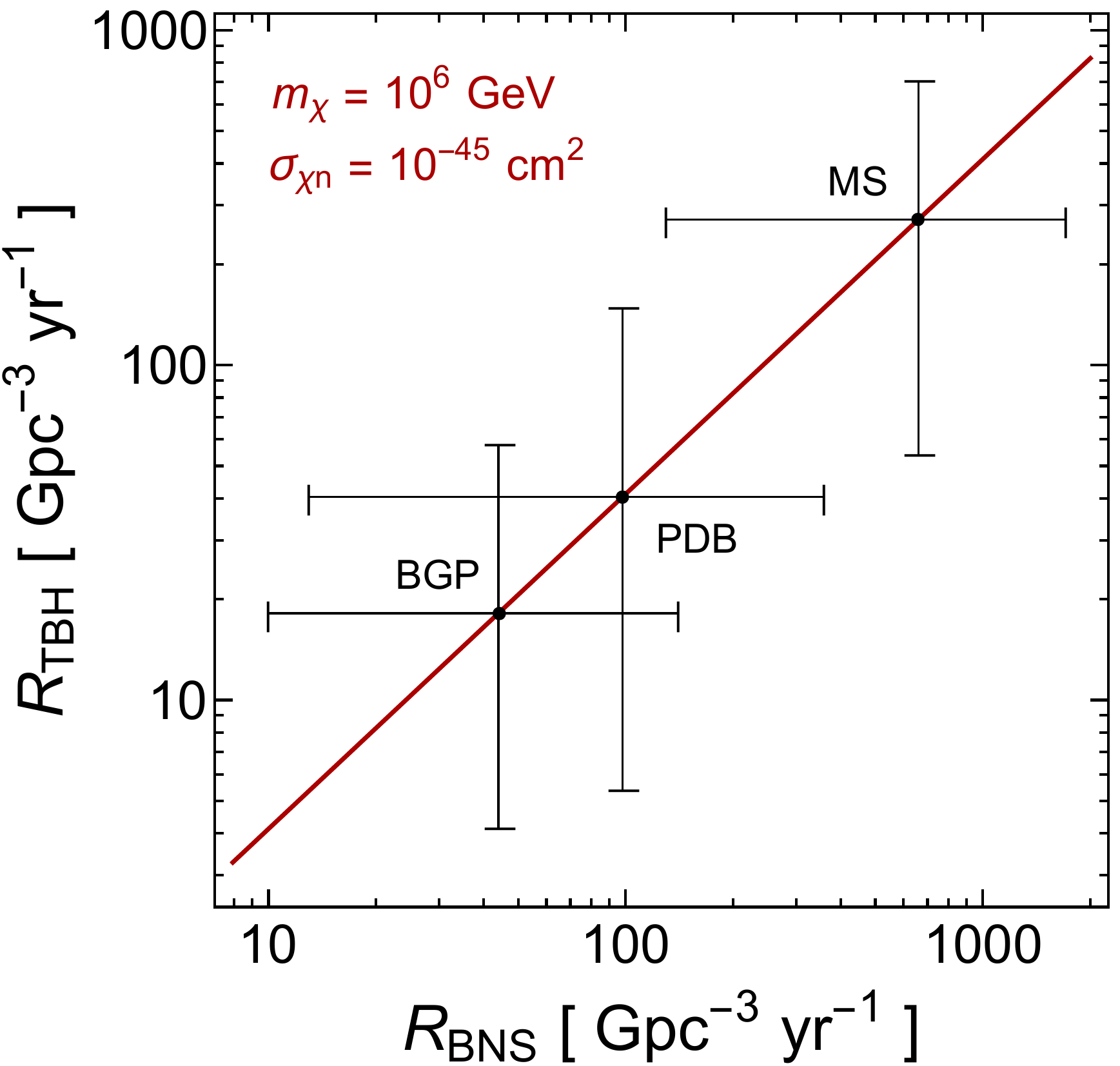}\qquad	\hspace{1.5 cm}
	\includegraphics[width=0.35\textwidth]{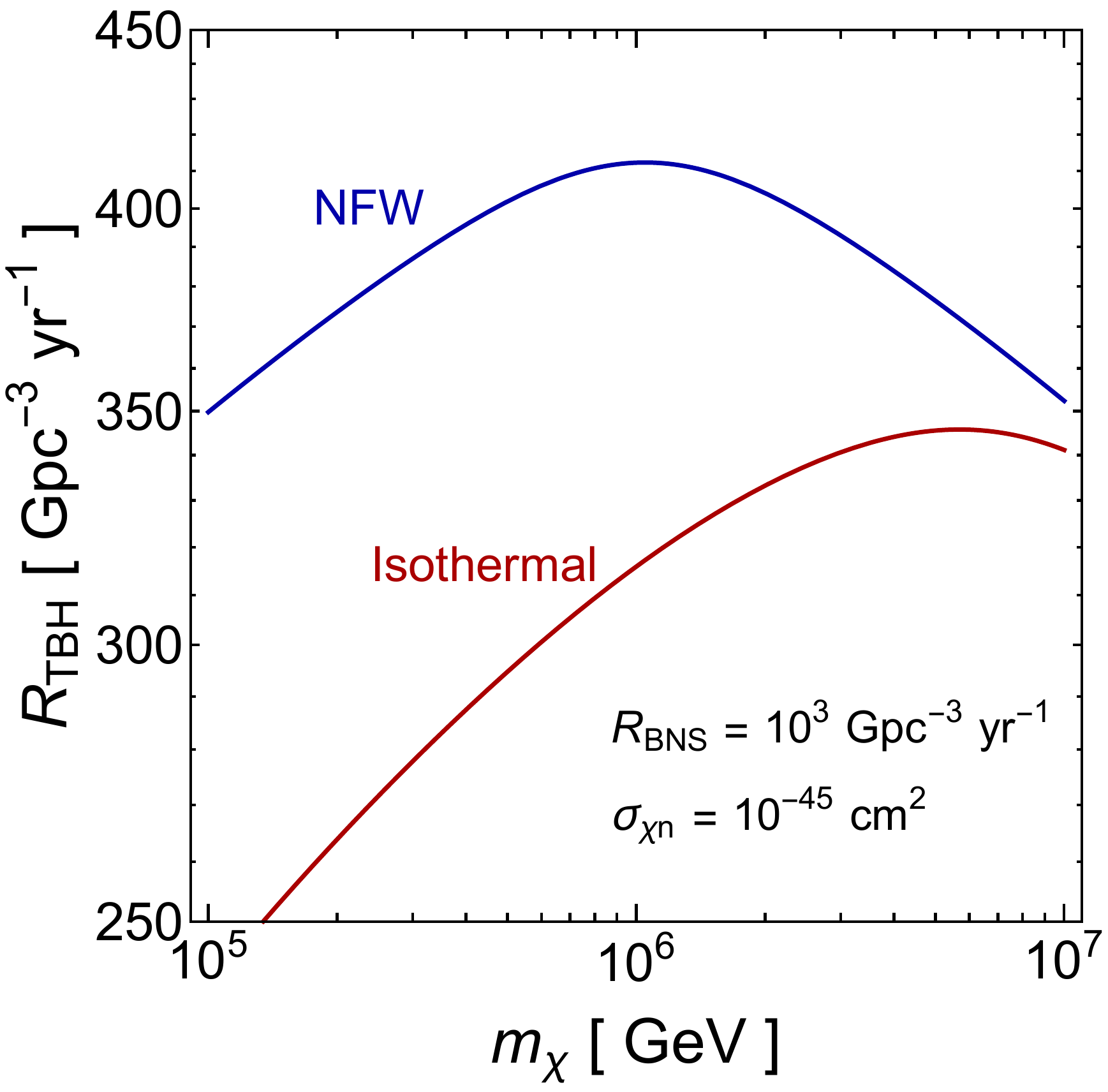}\\[2em]
	\includegraphics[width=0.35\textwidth]{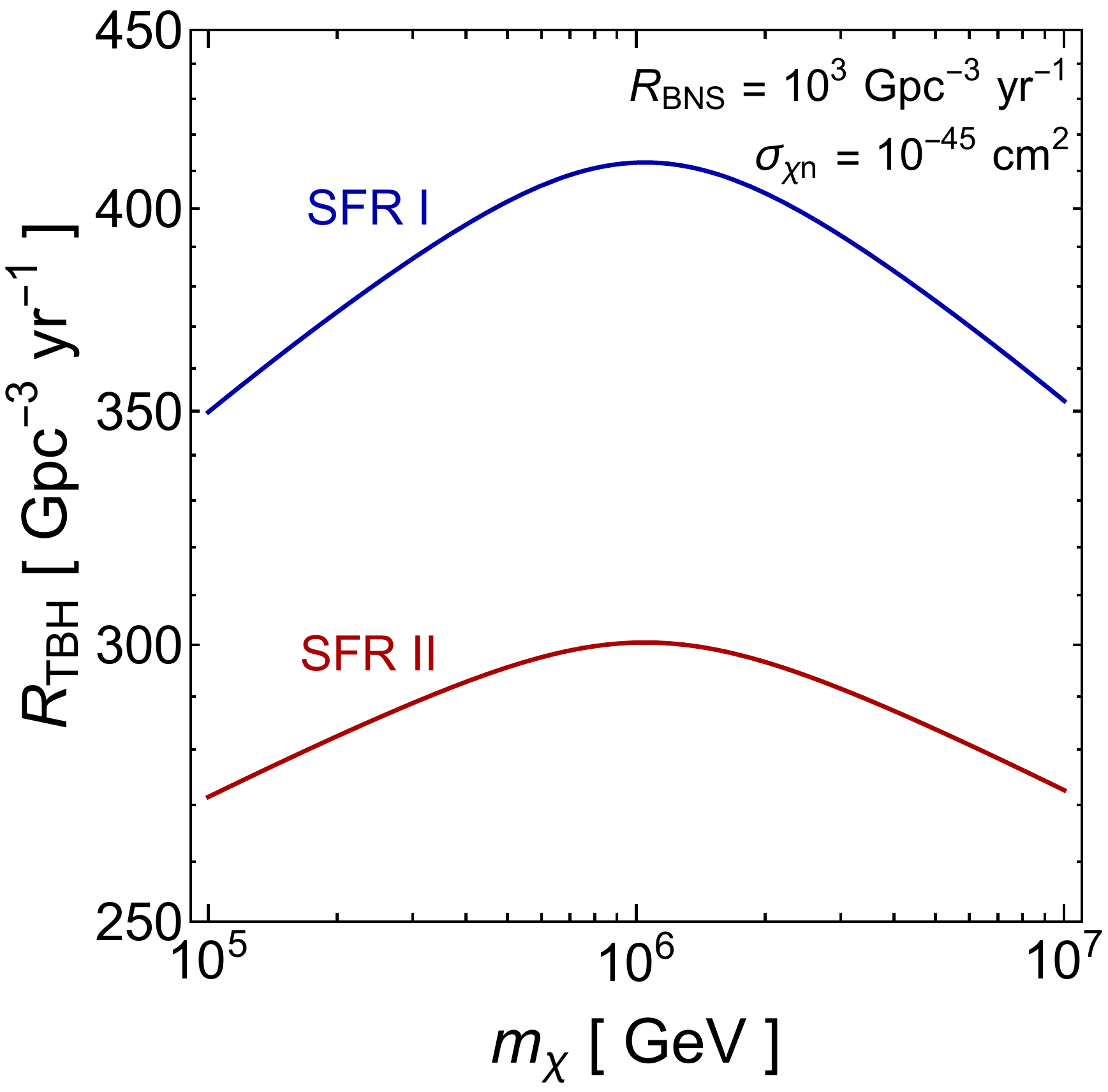}\qquad \hspace{1.5 cm}
	\includegraphics[width=0.35\textwidth]{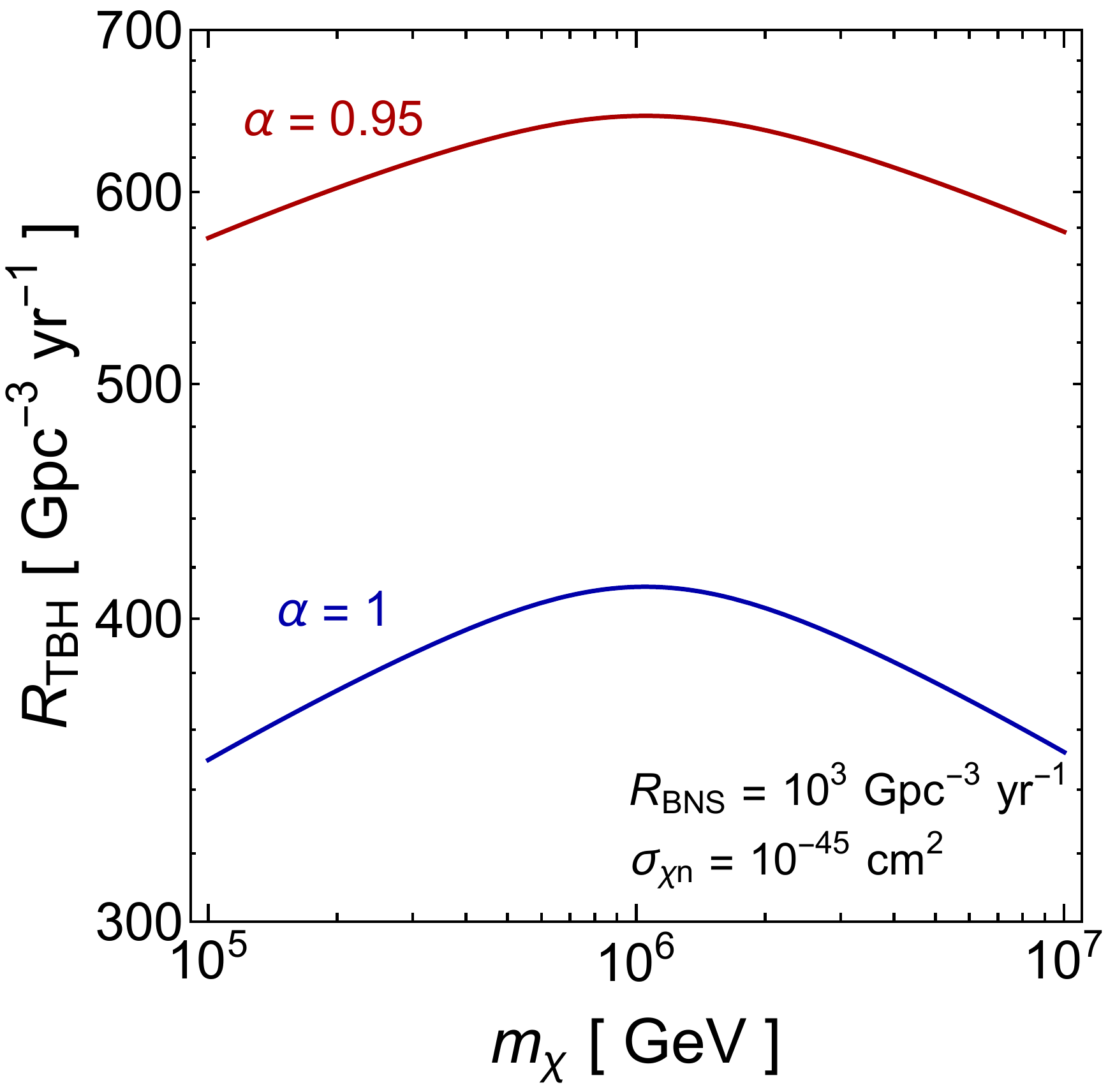}
	\caption[]{Variation of the TBH merger rate density with the astrophysical inputs, i.e., BNS merger rate density (top left panel), DM density profile (top right panel), star formation rate (bottom left panel), and merger time distribution of BNSs (bottom right panel). See Sec.\,\ref{sec:astroprop} for further discussion.}
	\label{dmdensity}
\end{figure*}

We explore how astrophysical uncertainties on the BNS merger rate, the DM density profile, the cosmic star formation rate model, and the delay time distributions can affect the TBH merger rate density.
\subsubsection{BNS Merger Rate}\label{bnsmerge}
BNS merger rate sets the normalization of $R_{\rm{TBH}}$.  Owing to recent observations of a few BNS mergers, LIGO provides an estimate of the BNS merger rate, which ranges from (10 $-$ 1700)\,Gpc$^{-3}$ yr$^{-1}$~\cite{LIGOScientific:2021psn}. As a result, the normalization of $R_{\rm{TBH}}$ is uncertain by two orders of magnitude. In Fig.\,\ref{dmdensity} (top left panel), we show the dependence of TBH merger rate density on $R_{\rm{BNS}}$, superimposing the $R_{\rm{BNS}}$ measurements for various population models for BNSs. At present, this turns out to be the largest source of uncertainty for the estimate of $R_{\rm TBH}$. For the present study, we will marginalize over the above-said allowed range of $R_{\rm{BNS}}$ to present our benchmark constraints and forecasts.

\subsubsection{DM Density Profiles}\label{dmdn}
In the main analysis, we assume a Navarro-Frenk-White profile~\cite{Navarro:1995iw,Navarro:1996gj} for the DM density distribution in the halos. Here, we explore a possible variation and consider a cored isothermal profile. Since, we are considering the progenitors in the inner parts of the galaxies, DM density distribution (cored/cuspy) plays a key role on the TBH merger rate density. We compute the TBH merger rate density for the cored isothermal profile, and in Fig.\,\ref{dmdensity}, we compare it with the fiducial Navarro-Frenk-White result. We use the following parameters for the cored isothermal profile: $\rho_{\rm{iso}}= \rho_0/\left(1+(r/r_0)^2\right)$ with $\rho_0=1.387$\,GeV/cm$^3$ and $r_0=4.38$\,kpc~\cite{Boudaud:2014qra,1980ApJS...44...73B}. It is clear that the lower core density of the density profile leads to a significant reduction of $R_{\rm TBH}$ by prolonging the collapse time. For sufficiently large $m_{\chi}$, the transmutation time is determined by the swallow time instead. In this regime, $R_{\rm{TBH}}$ becomes less dependent on the DM density.

\subsubsection{Cosmic Star Formation Rate}
\label{sec:CSFR}
The theoretical estimate of the TBH merger rate density depends on the cosmic star formation rate models. We use the Madau-Dickinson star formation rate model~\cite{Madau:2014bja} for our benchmark estimate. Here, we consider an alternative star formation rate model (Steidel et al.~\cite{Steidel:1998ja,Porciani:2000ag}) to quantify the uncertainty on $R_{\rm{TBH}}$. In Fig.\,\ref{dmdensity} (bottom left panel), we compare the two results, where SFR I denotes the Madau-Dickinson star formation rate, and SFR II denotes the star formation rate from refs.\,\cite{Steidel:1998ja,Porciani:2000ag}. We note that, cosmic star formation models act as a normalization of the TBH merger rate density, and changes $R_{\rm{TBH}}$ by at most 20\%.

\begin{figure*}[!t]
	\centering
	\includegraphics[width=0.4\textwidth]{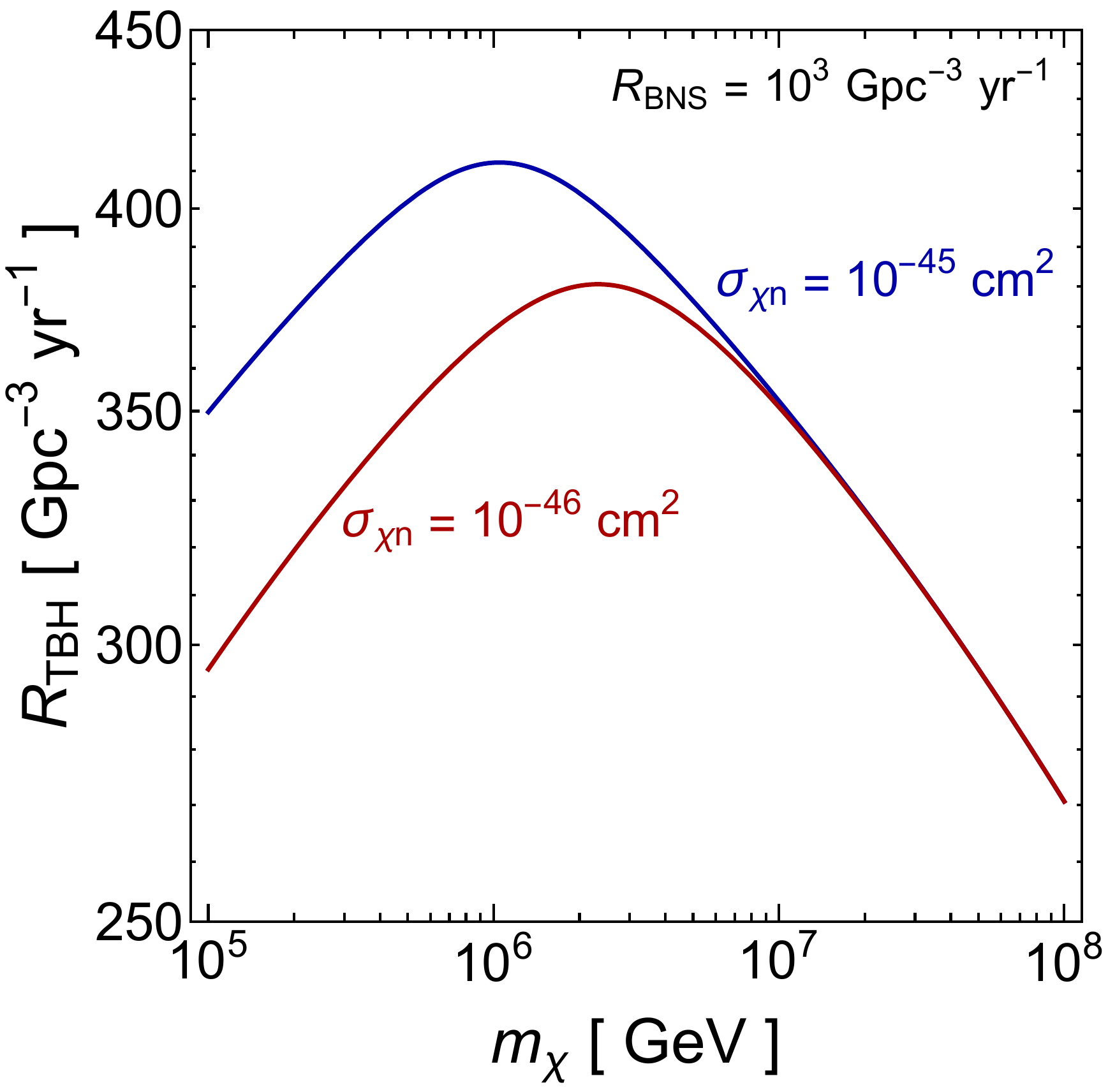}
	\hspace{0.9 cm}
	\includegraphics[width=0.4\textwidth]{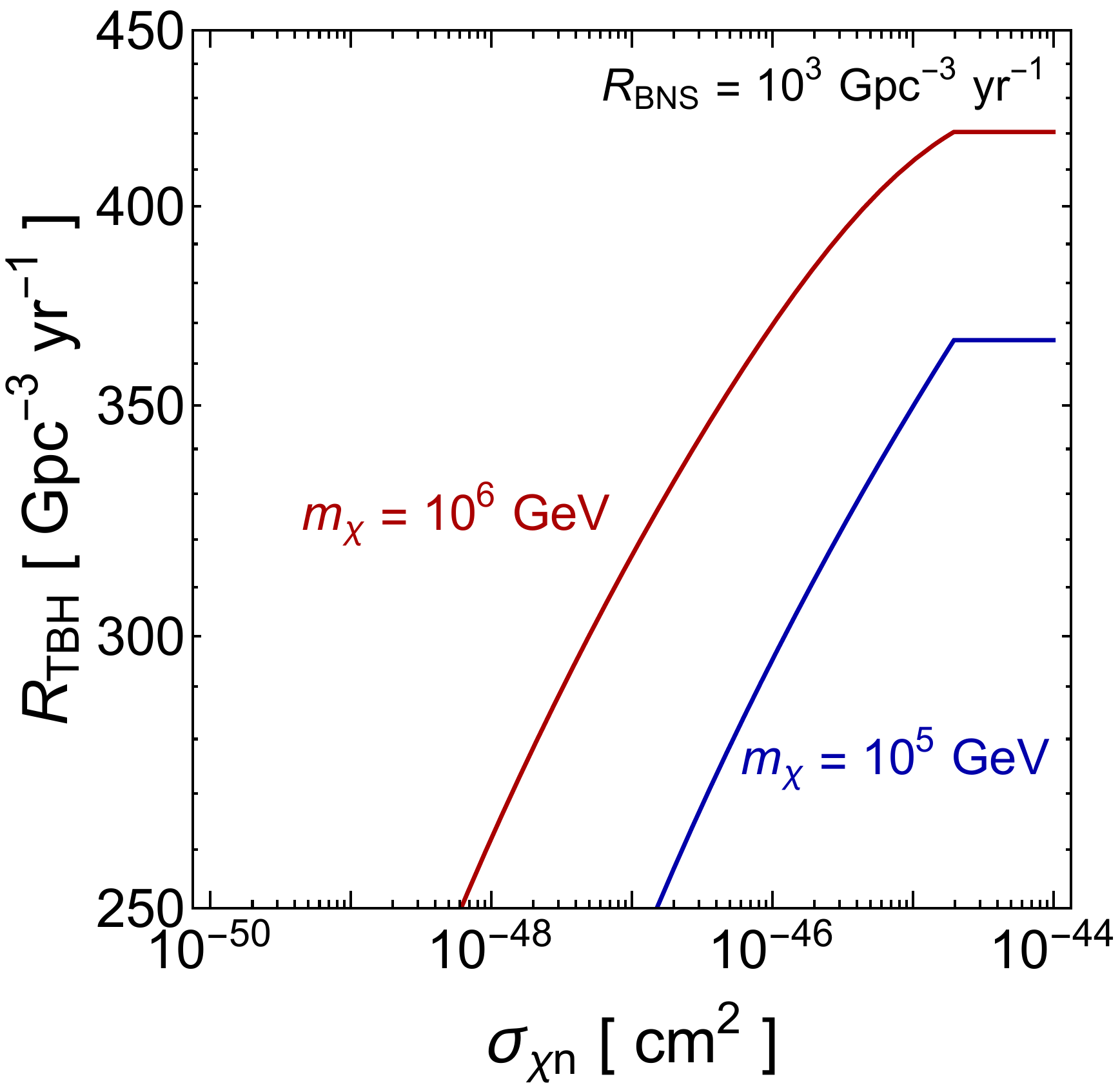}
	\caption[]{TBH merger rate density dependence on the DM mass (left panel) and DM-nucleon cross-section (right panel). See Sec.\,\ref{sec:DMprops} for further discussion.}
	\label{parameter}
\end{figure*}

\subsubsection{Delay Time Distribution}
\label{sec:DTD}
TBH merger rate density depends on the probability distribution of the delay time $(t_m-t_f)$,  i.e., the time interval between the formation and mergers of the binaries. For the main analysis, we consider a delay time distribution proportional to $1/(t_m-t_f)$, where $t_f$ denotes the binary formation time, and $t_m=t_0$ denotes the time of merger, which in our case is the current time. This choice is well-motivated by stellar population synthesis models~\cite{OShaughnessy:2009szr,2010MNRAS.402..371B,Dominik:2012kk}. However, because of the stellar metallicity, as well as uncertainties in the initial separation of the binaries, the delay time distribution can deviate from this form~\cite{Vitale:2018yhm,Santoliquido:2020axb,Mukherjee:2021qam}, and in order to bracket this uncertainty, we consider a generalized delay time distribution of $\sim 1/(t_0-t_f)^{\alpha}$ with $\alpha \in (0.5,1)$~\cite{Mukherjee:2021qam}. We found that the fiducial model chosen in our analysis leads to the most conservative result, and lower values of $\alpha$ always result in higher $R_{\rm{TBH}}$. In Fig.\,\ref{dmdensity} (bottom right panel), we compare the variation in $R_{\rm{TBH}}$ for $\alpha=0.95$, and we note that TBH merger rate density increases with decrease in $\alpha$. We also verify that exponential delay time distribution in~\cite{Vitale:2018yhm} leads to much higher $R_{\rm{TBH}}$ as compared to our fiducial model.

To summarize, astrophysical uncertainties such as DM density profiles, cosmic star formation rate models, and delay time distributions of the binaries  can affect the TBH merger rate density. Since the progenitors are mostly distributed in the inner part of the galaxies, cored/cuspy DM profiles have the most prominent impact on $R_{\rm{TBH}}$.  Whereas, star formation rate models and delay time distributions merely act as a normalization, and the impact of their uncertainty is almost insignificant compared to that of $R_{\rm BNS}$. We summarize the key assumptions on the progenitor and astrophysical properties in the TBH merger rate calculation and also their effect on $R_{\rm TBH}$ in Table\,\ref{table}.
\def\arraystretch{1.3}
\begin{table}[!h]
	\centering
	\begin{ruledtabular}
		\begin{tabular}{p{0.2cm}p{6.6cm}p{5.5cm}c}
			& \textbf{Parameters} & \textbf{Benchmark} & \textbf{Impact on $R_{\rm{TBH}}$} \\
			\hline
			& Progenitor mass (see Sec.\,\ref{sec:NSmass}) & 1.35 $M_{\odot}$  &  \makecell{\\ Increases by (12-15)\% \\ with $M_{\rm NS}= (1-2)\,M_{\odot}.$ \\\medskip See SM Fig.\,\ref{progenitor} (left panel).\\   \\}\\
			& Progenitor radius (see Sec.\,\ref{sec:NSrad})& 10 km & \makecell{Decreases by (12-15)\% \\with $R_{\rm NS}=(10-20)\,{\rm km}$.\\\medskip See SM Fig.\,\ref{progenitor} (middle panel).\\  \\} \\
			
			& Progenitor core temperature (see Sec.\,\ref{sec:NStemp}) & 2.1 $\times 10^6$\, K& \makecell{Decreases by (18-23)\% \\with $T_{\rm core}=10^6\,\rm K-10^7\,K$.\\\medskip See SM Fig.\,\ref{progenitor} (right panel).\\ \\} \\
			
			& DM density profile (see Sec.\,\ref{dmdn}) & Navarro-Frenk-White (NFW)~\cite{Navarro:1995iw,Navarro:1996gj}& \makecell{Decreases at most by $\sim$ 30\%\\for isothermal profile \cite{Boudaud:2014qra,1980ApJS...44...73B}. \\\medskip See SM Fig.\,\ref{dmdensity} (top right panel).\\ \\} \\
			
			& Delay time distribution (see Sec.\,\ref{sec:DTD}) & $\alpha$ = 1& \makecell{Increases by a factor of $\sim$ 2\\ for $\alpha=0.95$.\\\medskip See SM Fig.\,\ref{dmdensity} (bottom left panel).\\ \\} \\
			
			& Cosmic star formation rate (see Sec.\,\ref{sec:CSFR}) & Madau \& Dickinson~\cite{Madau:2014bja} & \makecell{Decreases 23\% \\ for Steidel et al. rate \cite{Steidel:1998ja,Porciani:2000ag}.\\ \medskip See SM Fig.\,\ref{dmdensity} (bottom right panel).\\ \\}\\
			
			& Normalization (see main text \& Sec.\,\ref{bnsmerge})  & $R_{\rm{BNS}} = (10-1700)\,\rm Gpc^{-3}\,yr^{-1}$& \makecell{Scales linearly.\\\medskip See SM Fig.\,\ref{dmdensity} (top left panel). }
		\end{tabular}
	\end{ruledtabular}
	
	\caption{Summary table for TBH merger rate inputs and its dependence on input parameters.}
	\label{table}
\end{table}

\subsection{DM Parameters}
\label{sec:DMprops}

In this section, we qualitatively discuss the dependence of TBH merger rate density on DM parameters, such as DM mass, DM-nucleon scattering cross-section, and DM annihilation rate.

\subsubsection{DM Mass}
Merger rate of TBHs has a non-trivial dependence on DM mass. TBH merger rate  density depends on $m_{\chi}$ via transmutation time $\tau_{\rm trans}$, which is a sum of two timescales. The first is the collapse time ($\tau_{\rm{collapse}}$), the time required to accumulate enough DM particles for ensuing a dark core collapse, and it scales as $1/\sigma_{\chi n} m^{3/2}_{\chi}$ for bosonic DM. The second is the swallow time $(\tau_{\rm{swallow}})$, the time required by the nascent BH to consume the host, and it depends only on the DM mass ($ \sim m^{3/2}_{\chi}$). For relatively lighter DM masses, collapse time determines the transmutation. As a result, with increase in $m_{\chi}$, transmutation time gets shorter, and leads to a higher $R_{\rm{TBH}}$. For heavy DM masses, swallow time determines the transmutation, and since it scales as $ \sim m^{3/2}_{\chi}$, with further increase in $m_{\chi}$, transmutation time gets longer, lowering the $R_{\rm{TBH}}$. Note that, since the swallow time does not depend on the DM-nucleon scattering cross-sections, TBH merger rate density becomes $\sigma_{\chi n}$ independent for sufficiently heavy DM masses $m_{\chi} \geq \mathcal{O}(10^7)\,\rm{GeV}$. In Fig.\,\ref{parameter} (left panel), we show the DM mass dependence of the TBH merger rate density. Similar arguments explain the dependence on mass for fermionic DM.

\subsubsection{DM-Nucleon Scattering Cross-section}
In the optically thin regime, capture rate of incoming DM particles scales linearly with DM-nucleon scattering cross-sections. As a result, higher $\sigma_{\chi n}$ leads to larger capture rate, and therefore, shorter transmutation time. So, it is evident that TBH merger rate density increases with the DM-nucleon scattering cross-sections, and  becomes constant when $\sigma_{\chi n}$ reaches its geometric saturation limit. Quantitatively, for a NS of mass 1.35\,$M_{\odot}$, and $R=10$\,km,  $R_{\rm{TBH}}$ becomes constant for $\sigma_{\chi n} \geq 2 \times 10^{-45}$\,cm$^2$. In Fig.\,\ref{parameter} (right panel), we show the $\sigma_{\chi n}$ dependence of the TBH merger rate density. 

\subsubsection{DM Annihilations and Other Extensions}
We have explored the formation of TBHs  via gradual accumulation of non-annihilating particle DM inside NSs. Now, we quantify the critical annihilation cross-section below which TBH formation criterion holds.
For annihilating DM, the number of captured DM particles  inside the NS follows
\begin{equation}
	\frac{dN_{\chi}[t]}{dt} = C - C_{\rm{ann}}\,N^2_{\chi}[t]\,,
\end{equation}
where $C$ denotes the capture rate, and $C_{\rm{ann}}= \langle \sigma_{a} v \rangle/V_{\rm{th}}$ denotes the annihilation rate with $\langle \sigma_{a} v \rangle$ is the thermally averaged DM annihilation cross-section and $V_{\rm{th}}$ is the thermalization volume. The total number of captured DM particles within the star's lifetime $(t_{\rm{age}})$ is given by: $N_{\chi}[t_{\rm{age}}]= \sqrt{C/C_{\rm{ann}}} \tanh\left[t_{\rm{age}}/\tau_{\rm{eq}}\right]$, where $\tau_{\rm{eq}}= 1/ \sqrt{C\,C_{\rm{ann}}}$ denotes the equilibration time-scale.

For $\tau_{\rm{eq}} \leq t_{\rm{age}}$, capture rate equilibrates with the annihilation rate, and  $N_{\chi}[t_{\rm{age}}]=\sqrt{C/C_{\rm{ann}}}$ is essentially determined by the DM annihilation cross-section. In this regime, transmutation does not occur as the accumulation is very low. Whereas, for $\tau_{\rm{eq}} > t_{\rm{age}}$, $\tanh\left[t_{\rm{age}}/\tau_{\rm{eq}}\right] \sim t_{\rm{age}}/\tau_{\rm{eq}}$, and we recover the familiar expression of $N_{\chi}[t_{\rm{age}}]= C\,t_{\rm{age}}$, which holds for non-annihilating DM. In this regime, annihilations do not prohibit transmutations. This corresponds to an annihilation cross-section of
\begin{align}
	&\langle \sigma_{a} v \rangle \leq 4.3 \times 10^{-51}\,\textrm{cm}^3\,\textrm{s}^{-1}\left(\tfrac{0.4\, \rm{GeV\, cm^{-3}}}{\rho_{\chi}}\right)  \left(\tfrac{10^5\, \textrm{GeV}}{m_{\chi}}\right)^{1/2}\left(\tfrac{10^{-45}\,\rm{cm}^2}{\sigma_{\chi n}}\right)\left(\tfrac{T_{\rm{core}}}{2.1 \times 10^6\,\rm{K}}\right)^{3/2}\left(\tfrac{1\,\rm{Gyr}}{t_{\rm{age}}}\right)^2\,.
\end{align}
Quantitatively, for typical NS parameters, with $m_{\chi}=10^5$\,GeV and $\sigma_{\chi n} = 10^{-45}$\,cm$^2$, successful transmutation occurs for  $\langle \sigma_{a} v \rangle \leq 10^{-51}\,\textrm{cm}^3\,\textrm{s}^{-1}$. Note that this is many orders of magnitude smaller than the thermal relic cross-section~\cite{Steigman:2012nb}. This reiterates that the DM candidates for which the TBH formation may be relevant, must be produced either as an asymmetry or via some other non-thermal means.

We neglect possible self-interactions among the DM particles in the main text, and  in the following,  we qualitatively discuss the effects of DM self-interactions on TBH  formation. 

DM self-interactions can give rise to an additional contribution to the total capture rate as well as modifies the black hole formation criterion. The additional contribution to the total capture rate, often known as self-capture, is insignificant  for neutron stars as they possess an extremely large baryonic density~\cite{Dasgupta:2020dik}. Quantitatively, for DM self-interaction strength of $\sigma_{\chi \chi}/m_{\chi} \sim 1$ cm$^2$/g, consistent with  the 
colliding galaxy clusters, self-capture is $10^{-11} \times$ baryonic capture for $m_{\chi} = 10^5$ GeV, and it reduces further with increase in DM mass, demonstrating the insignificance of self-capture to the total capture rate. On the other hand, self-interactions among the DM particles plays a crucial role  on  BH formation criterion. Specifically, for a quartic repulsive self-interactions, the Chandrasekhar limit becomes significantly higher, prohibiting the black hole formation inside the stellar core, and leads to a complete wash out of the constraints for strong self-interactions~\cite{Bell:2013xk}.

We have restricted this study to TBH-TBH
mergers from the transmutation of binary NSs. It is
also possible that two isolated NSs transmute and subsequently form a binary, causing an additional contribution to the TBH merger rate density which is independent of the BNS merger rate density. Conservatively, we neglect this additional contribution owing to its uncertain rate. TBH-NS binaries are less likely in our scenario because, given similar ambient DM densities we do not expect one NS in the binary to transmute while another doesn’t. However, it is possible that an isolated NS transmutes to a BH, and subsequently form binary to another NS
or with more massive ordinary BHs, leading to TBH-NS binary or TBH-BH binary. In the latter case, it can
even form extreme mass-ratio inspirals (EMRIs), which
are particularly interesting sources to be probed by the
planned GW detectors such as LISA and Einstein Telescope~\cite{Barsanti:2021ydd}. Leptophilic interactions of non-annihilating DM~\cite{Garani:2018kkd,Joglekar:2019vzy,Bell:2020lmm} is another potential direction which can be explored in a future study. 

\begin{figure}[!t]
	\centering
	\includegraphics[width=0.92 \columnwidth]{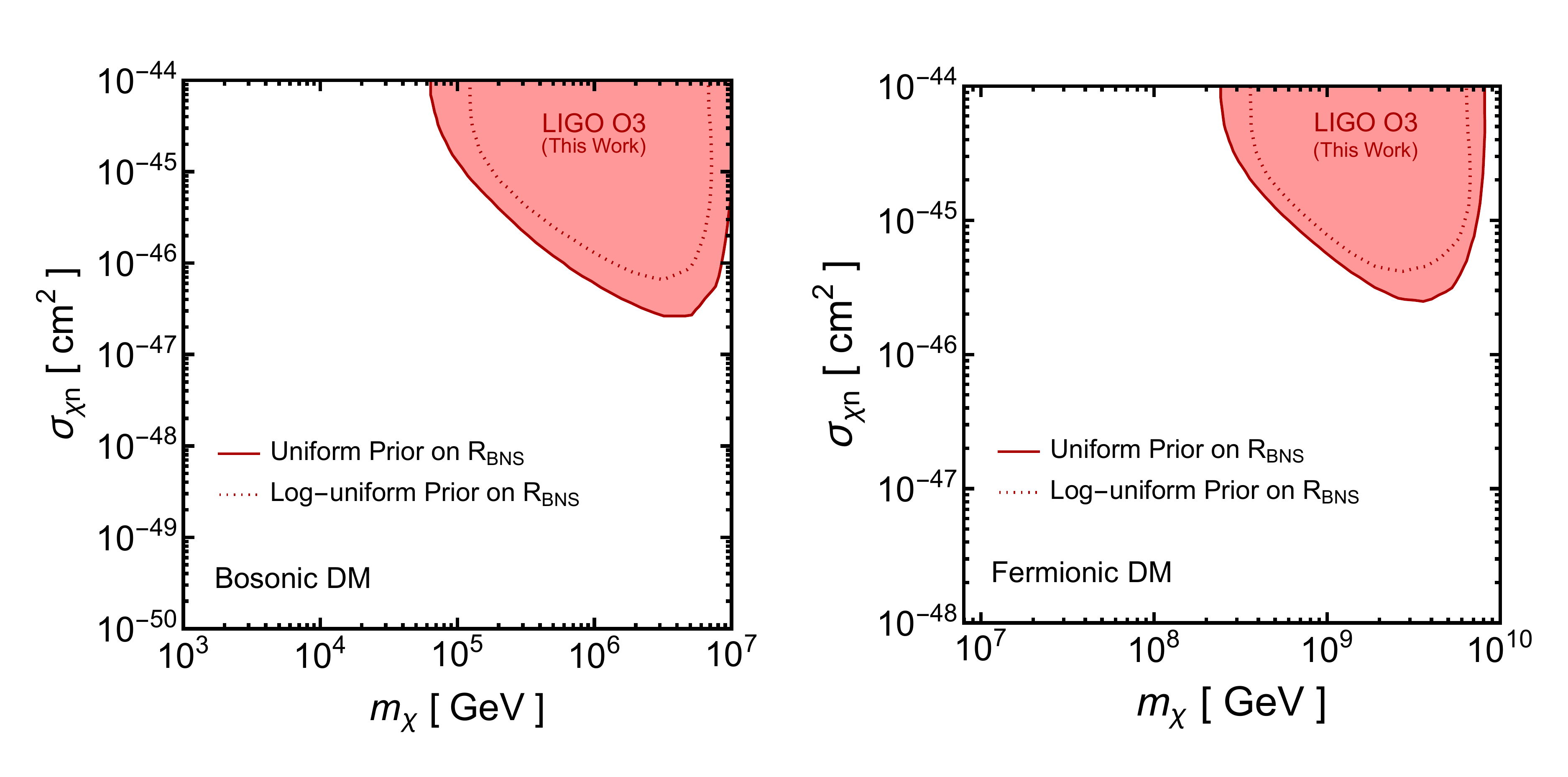}
	\caption{Impact of different choice of priors on the Bayesian exclusions. The left (right) panel is for non-annihilating bosonic (fermionic) DM. In each panel, region excluded by the solid line represents the \textit{uniform} prior on $R_{\rm BNS}\in\,$(10\,$-$\,1700)\,Gpc$^{-3}$yr$^{-1}$, whereas, region excluded by the dotted line represents the \textit{log-uniform} prior on $R_{\rm BNS}\in\,$(10\,$-$\,1700)\,Gpc$^{-3}$yr$^{-1}$. }
	\label{prior}
\end{figure}

It is also interesting to explore the potential improvements of the exclusion limits for next generation GW detectors. Quantitatively,  from null detection of $1M_{\odot}$ $-$ $1M_{\odot}$ compact binaries searches (with 1 year of observational time) the merger rate upper limits are estimated to be $R_{90} =  0.924\,(0.029)\, \rm Gpc^{-3}\,yr^{-1}$  for Einstein Telescope (Cosmic Explorer) \cite{Nunes:2021jsk,Maggiore:2019uih,Evans:2021gyd}, about $\sim 700\,(2 \times 10^4)$ smaller than the upper limits obtained by LIGO O3 data~\cite{LIGOScientific:2022hai}, indicating that our exclusion limits can putatively be improved by a factor of $\sim 700\,(2 \times 10^4)$ with next generation GW detectors, such as Einstein Telescope (Cosmic Explorer). Furthermore, with these next-generation GW detectors one can measure tidal deformability, allowing rejection of BNS events that could otherwise mimic BBH events.

\subsection{Dependence of priors on the Bayesian exclusion limits}
For the benchmark Bayesian analysis, we have considered log-uniform priors for the DM parameters (DM mass and DM-nucleon scattering cross-section) and a uniform prior for the BNS merger rate over the range of $\left[10-1700\right]\,\rm Gpc^{-3}\,yr^{-1}$. However, we have investigated the impact of different priors on our exclusion limits and found that the effect is at best modest. Quantitatively, by adopting a log-uniform prior for $R_{\rm{BNS}}$ in the same range, we obtain slightly weaker exclusion limits (at most by a factor of 2) as compared to the limits obtained for the uniform prior case, as shown in Fig.\,\ref{prior}.  This is simply because in the uniform prior case, the average value of $R_{\rm BNS}$ is 855 $\rm Gpc^{-3}\, yr^{-1}$, whereas, in the log-uniform case, the average value of $R_{\rm BNS}$ is $10^{0.5 \times \left(\log_{10}[1700]+\log_{10} [10]\right)}= 130\rm\, Gpc^{-3}\,yr^{-1}$. Since, the MCMC sampling favors relatively  lower values of $R_{\rm BNS}$ (maximal likelihood occurs for lower $R_{\rm BNS}$), we should expect that in the log-uniform prior case, the exclusion limits will only be slightly weakened.

\end{document}